\documentclass[10pt]{article}

\usepackage{amssymb}
\usepackage{amsfonts}
\usepackage{amsmath}
\usepackage[english, german]{babel}
\usepackage{latexsym}
\usepackage{color}
\usepackage[pdftex]{graphicx}
\usepackage{subfigure}

\newtheorem{thm}{Theorem}[section]
\newtheorem{proposition}[thm]{Proposition}
\newtheorem{corollary}[thm]{Corollary}
\newtheorem{lemma}[thm]{Lemma}
\newtheorem{definition}[thm]{Definition}
\newtheorem{remark}[thm]{Remark}

\newenvironment{Proof}{\textsc{Proof.}}{\mbox{ } \hfill $\Box$ \vspace{2mm}}

\newenvironment{rmenumerate}
    {\begin{enumerate}}
    {\end{enumerate}}

\def\dfrac{\displaystyle\frac}

\def\be{\begin{equation*}}
\def\ee{\end{equation*}}

\newcommand{\ba}{\begin{array}{ll}}
\newcommand{\bal}{\begin{array}{ll}}
\newcommand{\ea}{\end{array}}

\newcommand{\F}{\mathbb{F}}
\newcommand{\n}{\mathbb{N}}
\newcommand{\Prob}{\mathbb{P}}
\newcommand{\re}{\mathbb{R}}

\newcommand{\Lp}{\mathbb{L}}

\newcommand{\B}{{\mathcal{B}}}
\newcommand{\C}{{\mathcal{C}}}
\newcommand{\E}{\textnormal{E}}
\newcommand{\mL}{\mathcal{L}}
\newcommand{\mm}{\mathcal{M}}
\newcommand{\mP}{\mathcal{P}}
\newcommand{\V}{\textnormal{Var}}
\newcommand{\X}{{\mathcal{X}}}

\def\i{{\mbox{\rm{1}\hspace{-0.1in}\rm{1}\hspace{0.00in}}}}
\def\A{{\mbox{A}}}
\newcommand{\R}{\hat{R}}

\def\O{\Omega}
\def\o{\omega}
\def\t{\theta}
\def\T{\Theta}
\def\P{{{\textstyle{\mathbf{\Pi}}}}}

\begin{document}

\selectlanguage{english}

\renewcommand{\baselinestretch}{1.3}\normalsize


\title{Efficiency and Equilibria in Games of Optimal Derivative Design\thanks{We thank Guillaume Carlier,
Ivar Ekeland, and seminar participants at various institutions for
valuable comments and suggestions. We are grateful to Alexander
Fromm for his help in developing Section~\ref{sec:examples} of this
work. This paper was finalized while the authors were visiting the
Institute for Mathematical Sciences at the National University of
Singapore. Financial support from the Deutsche
Forschungsgemeinschaft through the SFB 649 "Economic Risk" and from
the Alexander von Humbold Foundation via a research fellowship is
gratefully acknowledged. }}

\author{\normalsize Ulrich Horst  \\[8pt]
        \small  Institut f\"{u}r Mathematik   \\
        \small  Humboldt-Universit\"{a}t zu Berlin \\
        \small  Unter den Linden 6\\
        \small  10099 Berlin \\
        \small  horst@math.hu-berlin.de
         \and
        \normalsize  Santiago Moreno--Bromberg \\[8pt]
        \small  Institut f\"{u}r Mathematik  \\
        \small  Humboldt-Universit\"{a}t zu Berlin \&\\
        \small  Univestit\"{a}t Z\"{u}rich\\
        \small  Institut f\"{u}r Banking und Finance \\
        \small  santiago.moreno@bf.uzh.ch
\vspace*{0.1cm}}

\maketitle

\vspace{-.9cm}
\begin{abstract}
In this paper the problem of optimal derivative design, profit maximization and risk minimization
under adverse selection when multiple agencies compete for the business of a continuum of heterogenous
agents is studied. The presence of ties in the agents' best--response correspondences yields discontinuous
payoff functions for the agencies. These discontinuities are dealt with via efficient tie--breaking rules.
In a first step, the model presented by Carlier, Ekeland \& Touzi (2007) of optimal derivative design by profit--maximizing agencies is extended to a multiple--firm setting, and results of Page \& Monteiro (2003, 2007, 2008) are used to prove the existence of (mixed--strategies) Nash equilibria. On a second
step we consider the more complex case of risk minimizing firms. Here the concept of socially efficient allocations is introduced, and existence of the latter is proved. It is also shown that in the particular case of the entropic risk measure, there exists an efficient ``fix--mix'' tie--breaking rule, in which case firms share the whole market over given proportions.

\end{abstract}

\vspace{2mm}




\hfill

\newpage

\renewcommand{\baselinestretch}{1.1}\normalsize

\section{Introduction}

This work lies at the intersection of the fields of multi--principal--multi--agent games under adverse selection
and optimal risk sharing. The former emerged as an extension into oligopolistic competition of the
Principal--Agent models already studied in the 1970's and early 1980's by (among others) Maskin, Mussa,
Rosen and Akerlof; whereas the latter can be traced back to Borch (1962) and Arrow (1963), and it has
experienced a rebirth of sorts with the introduction of the notion of convex risk measures in the late 1990's.

In the (standard) Principal--Agent model of non--linear pricing of hedonic goods, a monopolist (the principal) has the capacity to deliver
quality--differentiated products, which are assumed to lie on some
compact and convex set $C\subset\re^n.$ Such products are the {\sl
technologically feasible} goods from the point of view of the
principal. These vectors are usually called {\sl consumption
bundles}, and each of its coordinates indicates how much a certain
product possesses of a given attribute. In other words, goods are
assumed to be fully described by a list of qualities that are
relevant to the consumers. The buyers (or agents) whom the principal
engages with the intent to trade have heterogenous preferences.
This is captured by indexing the different characteristics or
agent {\sl types} with vectors in some set $\T\subset\re^m,$ and
including the types in the arguments of the buyers' utility
functions. The types are {\sl private information}, i.e. the
principal is aware of their statistical distribution $\mu,$ but
she cannot distinguish an agent's type prior to engaging him. The
ill--informed principal designs {\sl incentive compatible}
catalogues with the intention of (at least partially) screening
the market. The trading is done on a take--it--or--leave--it basis
and it is assumed there is no second--hand market. The
characterization of the solutions to problems of this kind can be
found, among others, in the works of Armstrong~\cite{kn:A}, Mussa
\& Rosen~\cite{kn:mr} and Rochet \& Chon\'e~\cite{kn:rch}.

The standard Principal--Agent model was extended by Carlier, Ekeland
\& Touzi in~\cite{kn:cet} to model a problem of optimal
(over--the--counter) derivative design. They assume there is a
direct cost to the principal when she delivers a derivative contract
or financial product (a typical example are mutual funds) and that
the agents' utilities are of mean--variance type {\footnote{These
are type-dependent utility functions of the form $U(\t,
X)=G(X)+\t\cdot F(X),$ where the asset--space is $\X,$ $G:\X\to\re$
is linear and $F:\X\to\T.$}}. This allows them to phrase the
principal's profit maximization problem as a problem in the Calculus
of Variations subject to convexity constraints, the latter capturing
the incentive compatibility constraints on the set of admissible
catalogues. This approach was further modified by Horst \&
Moreno--Bromberg in~\cite{kn:hm}. Here the authors assume the
principal is exposed to some non--headgeable risky position, which
she evaluates using a convex risk measure. This could be, for
example, an insurer who has issued claims that are correlated to
weather phenomena. The principal's aim is to minimize her exposure
by laying off part of her risk with heterogenous agents. To do so
she proceeds as in~\cite{kn:cet} and designs a catalogue of
derivatives written on her income, as well as a non--linear pricing
schedule. In contrast with Carlier, Ekeland \& Touzi, in this case
the impact of each individual trade on the principal's evaluation of
her risk is non--linear, and an individual trade does not
necessarily reduce the principal's exposure. The main results of
both papers are existence and characterization of direct revelation
mechanisms that maximize the principal's income (\cite{kn:cet}) or
minimize the principal's risk assessment (\cite{kn:hm}).

A more complex scenario, in which our current model is embedded,
contemplates an oligopoly. Analyzing oligopolistic competition,
even in the absence of adverse selection, is qualitatively more
daunting than dealing with single--seller settings. For example,
d'Aspermont \& dos Santos Ferreira study in~\cite{kn:DADSF} the
existence of (Nash) equilibria in a common agency game (no
asymmetric information). The presence of competition, and the
corresponding constraints that it adds to the firms' optimization
problems separates their model from single--principal settings.
They use Lagrange--multipliers methods to characterize the sets of
equilibria. Interestingly, the authors show that contingent on
different choices of the model's parameters, equilibrium outcomes
may range from fully competitive to collusive. Not surprisingly,
once one adds the adverse selection ingredient to the problem
things become more complex. The first challenge that needs to be
addressed when one studies existence of equilibria and/or Pareto
optimal allocations is the fact that the Revelation Principle, an
important simplifying ingredient in the Principal--Agent
literature, may no longer be applied without loss of generality.
In other words, some equilibrium allocations may not be
implementable via direct revelation mechanisms. There are two ways to overcome this. The
first one, introduced by Epstein \& Peters~\cite{kn:ep}, consists
in enlarging the message space to include not only the agent types,
but also a description of the market situations. Despite the
theoretical value of this result, it is in general not practical for
applications. Alternatively, Martimort \& Stole ~\cite{kn:MaSt} and
Page Jr. \& Monteiro~\cite{kn:PaMo2} developed a multi--agency
analogue of the Revelation Principle, the Delegation Principle,
which allows enough simplification of the general non--linear
pricing game to have a workable setting. A second challenge is the
emergence of ties in the agents' best--response mappings. When
agents are indifferent between contracting with different firms,
then these firms' payoff functions may have discontinuities. This
precludes the use of the classical results of Debreu, Glicksberg and
Fan (see for example~\cite{kn:ft}) to guarantee the existence of
Nash equilibria. Instead, much of the current literature on the
existence of Nash equilibria in multi--firm--agent games relies on
Reny's seminal paper~\cite{kn:Re}, and the necessary conditions he
presents for the existence of Nash equilibria in discontinuous
games. The results of Bagh \& Jofr\'e in~\cite{kn:BaJo}, and of Page Jr. \& Monteiro in~\cite{kn:PaMo3}
and~\cite{kn:PM4}  provide
testable conditions that allow for Reny's results to be used in
multi--firm--agent games. An alternate route has been taken by
Carmona \& Fajardo in~\cite{kn:CaFa}. The authors provide an
existence result of sub--game perfect equilibria in common agency
games. Just as Martimort \& Stole and Page \& Monteiro have done,
they concentrate on catalogue games. They, however, present an
extension of Simon \& Zame's theorem that allows them to relax the
assumption on exclusivity in contracting found, among others,
in~\cite{kn:PaMo2}. Moreover, their model exhibits intrinsically
generated sharing rules. The price to pay for this level of
generality is the required assumption of continuous (instead of
upper semicontinuous) payoff functions of the firms.

A second field that influenced this work could be broadly labeled
``risk minimization and sharing'' with coherent or, more generally,
convex risk measures. The notion of {\sl coherent risk measures} was
axiomatized by Artzner, Delbaen, Ebert \& Heath in~\cite{kn:ADEH} as
an ``acceptable'' way to assess the riskiness of a financial
position. It was then extended to {\sl convex risk measures} by,
among others, F\"{o}llmer \& Schied~\cite{kn:fs2} and Frittelli \&
Rosazza Gianin~\cite{kn:FRG}. This theory has experienced an
accelerated development, as evidenced by a large number of
publications. Some quite natural questions to address, in a
multitude of settings, are the existence and structure of
risk--minimizing positions or of risk--sharing allocations that are
either Pareto optimal or that constitute an equilibrium of some
sort. The problem of optimal risk sharing for convex risk measures
was first studied by Barrieu \& El~Karoui in~\cite{kn:BK3}. They
gave sufficient conditions for the risk--sharing problem with
general state spaces to have a solution. Their conditions can be
verified for the special case where the initial endowments are
deterministic and the agents use modifications of the same risk
measure; Jouini, Schachermayer \& Touzi proved in~\cite{kn:jst} the
existence of optimal risk--sharing allocations when the economic
agents assess risk using convex risk measures which are {\sl
law--invariant}. The optimal allocations are Pareto optimal but not
necessarily {\sl individually rational}. That is, a cash transfer,
also called the {\sl rent of risk exchange}, could be necessary to
guarantee that the outcome leaves all parties better off than they
originally were. It should also be mentioned that the setting in
in~\cite{kn:jst} is over--the--counter (OTC) in nature, hence the
question of efficiency and implementability cannot be left to the
market. Implementability could depend on the presence of a social
planer who enforces cash--transfer schemes or other policies that
generate individually rational or socially optimal outcomes. In
contrast, equilibrium models do not require the presence of a
regulator to guarantee the implementability of efficient
allocations. Implementability and efficiency is left up to the
(financial) market. Equilibrium models of incomplete markets where
agents use convex risk measures to evaluate their risk exposures
were studied by, e.g., Filipovi\'c \& Kupper in~\cite{kn:FK1} in the
static case and Cheridito et al. \cite{kn:chkp} in the dynamic one.

In this paper we study the problem of optimal derivative design, profit
maximization and risk minimization under adverse selection when multiple
agencies compete for the business of a continuum of heterogenous
agents. We first extend the model of profit maximization presented
in~\cite{kn:cet} to a multi--firm one. Here is where the theory of
games played under adverse selection comes into play. In order to
use recent results from the theory of catalogue games, we assume
that each firm's strategy set consists of the closed, convex hull
of a finite number of basic products. We believe this is
consistent with the idea that each firm is exposed to a direct
cost when delivering financial products or derivative securities, since it must
purchase the underlying assets that shall be structured into its'
product line. In mathematical terms, such assumption renders the
strategy sets compact with respect to the same topology for which
the players' preferences are continuous (a necessary assumption),
and therefore makes the game in hand compact. We show that when
the game is played under a particular kind of tie--breaking rules,
then it is uniformly payoff secure and reciprocally upper
semicontinuous and hence has a mixed--strategies Nash equilibrium.
Such rules, which paraphrasing~\cite{kn:PM4} we call {\sl
efficient}, might not be implemented if not for the presence of a
social planer. The regulator also plays a key role in our
multi--agency extension of the risk minimization model studied
in~\cite{kn:hm}. In both cases the impact of a single trade on the
firm's risk evaluation is highly non--linear and depends strongly
on the firm's overall position.\footnote{In a model of profit
maximization a firm contracts with some agent type {\sl
independently} of other types as long as that particular type
contributes positively to the firm's revenues.} This results in
two considerable technical difficulties. First, there is no reason
to expect (and in general it will not be the case) that the firms'
payoff functions will be uniformly payoff secure.\footnote{The
``worst--case'' tie--breaking rule, where each firm {\sl assumes}
ties will be broken in the most disadvantageous way for itself, does yield
payoff security, but at the cost of reciprocally
upper  semicontinuity.} Second, most of the current results
(\cite{kn:BaJo}, \cite{kn:ft}, \cite{kn:PaMo2},
\cite{kn:NR},...) on existence of Nash equilibria require some
form of (quasi-) concavity (except~\cite{kn:T} where a complete,
yet very challenging to verify, characterization of Nash
equilibria is presented). In our model these conditions are only
satisfied by the mixed extension of the game. Even if the game
were uniformly payoff secure and (weakly) reciprocally upper
semicontinuous (by no means a given),\footnote{This would imply
the existence of Nash equilibria in mixed strategies} considering
a mixed extension of the game would mean the following: firms view
linear aggregation of possible risk evaluations as the way to
assess the influence of others' in their own risk. Whether this
approach is consistent with the ideas behind the theory of convex
risk measure is in our opinion debatable. Taking the previous
arguments into account, we do not seek the existence of
(mixed--strategies) Nash equilibria. Our focus is instead on an
existence proof for {\sl socially efficient allocations} of risk
exposures. Such risk allocations minimize the firms' aggregate
risk and can hence be thought of as the multi--firm--agent game
analogous of the Pareto optimal allocations described
in~\cite{kn:jst}. The proof of existence of efficient
tie--breaking rules relies heavily on the fact that for fixed
price schedules and tie--breaking rule, the contracts that
minimize the aggregate risk can be expressed as the product of a
type--independent random variable and a coefficient that depends
exclusively on types. This separation result was already observed
in~\cite{kn:hm}, but it is in this paper where it is truly
exploited.
It is important to mention that efficient tie--breaking
rules are generated endogenously. This implies that in our OTC
model efficiency and regulation go hand--in--hand, even in the
absence of a cash--transfer scheme. When it comes to implementability, we have obtained
for the case of the entropic risk measure that among all efficient tie--breaking--rules, there is a constant ratio of market shares. In other words, there is a ``fix--mix'' ratio under which firms share the whole market, rather than segmenting it by agent types. The latter implies that all the firms offer the same indirect utility to each of the consumers. A real--world example where firms offer consumers essentially the same utility (using different products), and where the assumption of mean--variance optimizing consumers is appropriate, is retail banking. In such case ``fix--mix'' is also a reasonable assumption: retail banks differentiate customers according to their risk aversion when designing portfolio strategies, but do not necessarily try to appeal differently to customers with different attitudes towards risk. The methodology we used to obtain the ``fix--mix'' result suggests that it carries over to more general risk measures. A full analysis would be quite technical though, and certainly beyond the scope of this paper. We illustrate this result with a numerical example, in which we also find that (as expected) firms are worse off in the presence of competition, whereas the contrary is the case for the buyers. Moreover, the aggregate risk in the economy is better dealt with in the competitive, yet regulated, case. The numerical algorithm that we use for this example, a hybrid descent method, is also used to analyze an example where the firms are AV@R--minimizers. We find quite a sharp contrast  in the firms' risk profiles before and after trading when comparing the entropic and AV@R cases, which obeys the fact that when minimizing the latter, one should focus in the worst state of the world (our examples use, unavoidably, a finite probability space) in as much as the problem's constraint allow, then move to the second worse and so on.

The remainder of this paper is organized as follows:
Section~\ref{section:Gen_Fram} contains the description of our
general framework. We leave the setting as general as possible
while still describing the asymmetry of information in the model,
the best--response sets of the agents that give rise to ties,
tie--breaking rules and the influence of the social planer. In
Section~\ref{sec:ProfitMax} we study the game played among
profit--maximizing firms, where the main result is the existence
of mixed--strategies Nash equilibria. Section~\ref{sec:RiskMin} is
devoted to the risk--minimization game, where we prove the
existence of socially efficient allocations. Finally, we present in Section~\ref{sec:examples}
numerical algorithms to estimate equilibrium points and socially efficient allocations in some
particular examples, as well as the ``fix--mix'' result mentioned above.

\section{General Framework}\label{section:Gen_Fram}

We consider an economy that consists of two {\sl firms} and a
continuum of {\sl agents}. We study both the case when the firms are
profit maximizers, and the one where their objective is to minimize
the risk assessments of some initial uncertain payoffs. The analysis
of the two scenarios require different mathematical techniques and
distinct notions of efficient allocations, so we study the
two models separately.\footnote{Our arguments can be extended to an
economy with $m$ firms. We chose to work on the case $m=2$ for
simplicity.}

\subsection{The financially feasible sets}

The firms compete for the agents' business by offering derivatives
contracts. The set from which firm $i=1,2$ may choose
products, i.e. its {\sl financially feasible} set, is
\be
    \X_i\subset\Lp^2\left(\Omega,{\cal{F}}, \Prob \right),
\ee
where $\left(\Omega,{\cal{F}}, \Prob \right)$ is a standard probability space. We assume these sets are closed, convex and bounded, and that
$0\in\X_i.$ The boundedness of $\X_i$ implies there is $M>0$ such
that $\|X_i\|_2\le M$ for all $X_i\in\X_i.$ Any additional
requirements on the sets $\X_i$ shall be introduced when
necessary.\footnote{We omit writing $i=1, 2$ when we refer to
properties shared by both firms.} Throughout this paper we use the
notation
\be
    X_n\xrightarrow{\|\cdot\|_2}X,\quad X_n\xrightarrow{a.s.}X\quad\textnormal{and}\quad X_n\xrightarrow{w.}X
\ee
to indicate that the sequence $\{X_n\}$ converges strongly, almost surely and weakly to $X.$ We also use $\i_A$ to denote the indicator function of a set $A.$

In the literature on multi--firm, non--linear pricing games, it is
generally assumed that each firm chooses a compact subset
${\mathcal Y}_i$ from its financially feasible set, and it devises
a (non-linear) pricing schedule
\be
    p_i:{\mathcal Y}_i\to\re .
\ee
Each pair $(X, p_i(X))$ (where $X\in\X_i$) is called a {\sl
contract}. Then the Delegation and Competitive Taxation Principles
(see for example~\cite{kn:PaMo2}) allow for a
without--loss--of--generality analysis of the existence of Nash
equilibria by studying the (simpler) catalogue game played over
compact subsets of the product--price space $\X_i\times\re.$ In fact, the boundedness
assumption on $\X_i$ implies that prices belong to some compact set
$P\subset\re.$ We also study games played over catalogues, but their structure will be
case--dependent. We write $C_i$ to denote a catalogue offered by
firm $i,$ and $(C_1, C_2)$ for a {\sl catalogue profile} (also
called a market situation). We postpone the specification of the
criterions that the firms aim to optimize until Sections
\ref{sec:ProfitMax} and \ref{sec:RiskMin}.

\subsection{The agents' preferences}\label{sec:Agents}

The  agents are heterogenous, mean--variance maximizers whose set of types (or characteristics) is
\be
    \Theta = [a,1] \quad\mbox{for some }\quad a > 0.
\ee
The right endpoint of $\Theta$ has been normalized to $1,$ but it
could be any finite value. What is required is for the set of
types to be a compact subset of the strictly positive real numbers.
An agent's type represents her risk aversion (hence the assumption $a>0$ means there are no risk--neutral agents). In other words, given a contingent claim $Y$ an agent
of type $\theta$ assesses its worthiness via the (type--dependent) utility function
\be
    U(\t, Y)=\E[Y]-\t\, \V [Y].
\ee
The types are private information and not transparent to the
firms. They are distributed according to a measure $\mu$, which we
assume is absolutely continuous with respect to Lebesgue measure.
The measure $\mu$ is known to all firms. In other words, firms
cannot distinguish an agent's type when engaging in trading with her
(or they are legally prevented from doing so) but they know the
overall distribution of types. This asymmetry of information, also
known as {\sl adverse selection}, prevents the firms from extracting
all the above--reservation--utility wealth form each agent. The
knowledge of $\mu$ is therefore essential for the
agencies.\footnote{We could make additional assumptions on $\mu$ and
allow for $a=0.$} In the upcoming sections we require the following auxiliary lemma:

\begin{lemma}\label{lemma:equicont} The family of functions ${\cal{U}}=\{U(\t,\cdot)\,\mid\,\t\in\T\}$ is uniformly equicontinuous.
\end{lemma}
\begin{Proof}
Let $\epsilon>0$ and consider $X, X'\in\X,$ and $\t\in\T,$ then
\be
|U(\t,X)-U(\t, X')|=|\E[X]-\t\, \V[X]-\E[X']+\t\, \V[X']|.
\ee
From the triangle inequality and the fact that $0<\t\le 1$ we get
\be
|U(\t,X)-U(\t, X')|\le |\E[X]-\E[X']|+ |\V[X]-\V[X']|.
\ee
It follows from the Cauchy-Schwartz inequality that
\begin{small}
\be
|\E[X]-\E[X']|\le d_{\Prob}(X, X')\,\,{\textnormal{and}}\,\,|\V[X]-\V[X']|\le d_{\Prob}(X, X')\left(\int_{\O}|X+X'|d\Prob+|\E[X]+\E[X']|\right),
\ee
\end{small}
where $d_{\Prob}(X, X')$ is the $L^1$--distance with respect to $\Prob.$ Since $-M\le X, X'\le M,$ we obtain
\be
|U(\t,X)-U(\t, X')|\le d_{\Prob}(X, X')\left(1+4\,M\right)\le \left(1+4\,M\right)\|X - X'\|_2.
\ee
Setting $\delta=\epsilon\,\left(1+4\,M\right)^{-1}$ yields the desired result.

\end{Proof}
\subsection{Indirect utilities and best--response sets}\label{subsec:IndU_BestR}

When an agent faces a catalogue profile $(C_1, C_2),$ she chooses
a single contract from some $C_i$ on a take--it--or--leave--it basis.
Agents may choose any (available) contract they wish, but no bargaining regarding prices or
products takes place. Given $(C_1, C_2),$ the indirect utility of an agent of type $\t$
is:
\be
    v(\t, C_1, C_2)\triangleq\sup\big\{U(\t, X)-p\,\mid\, (X,p)\in
    C_i\,\,{\mbox{for some}}\,\,i\big\}.
\ee
We assume that the agents have an outside option that
yields their reservation utility, which we normalize to zero for all
agents. For a fixed market situation, the function $v(\cdot, C_1,
C_2):\T\to\re$ is convex, since it is defined as the pointwise
supremum of affine functions of $\t.$ The presence the outside
option guarantees $v\ge 0.$ The best--response set of the agents of type $\t$ to a
certain catalogue profile is defined as:
\be
    \mL(\t, C_1, C_2)\triangleq\textnormal{argsup}\big\{U(\t,
    X)-p\,\mid\, (X,p)\in C_i\,\,{\mbox{for some}}\,\,i\big\}.
\ee
In Sections~\ref{sec:ProfitMax} and \ref{sec:RiskMin} we make necessary assumptions on $C_i$ as to
guarantee $\mL(\t, C_1, C_2)\neq\emptyset$ for all $\t\in\T.$
Agents of type $\t$ are indifferent among the elements of $\mL(\t, C_1
,C_2),$ which they strictly prefer over any other contract that
can be chosen from $(C_1, C_2).$ The Envelope Theorem (see for
example~\cite{kn:ms}) implies that for any $(X(\t),
p(\t))\in\mL(\t, C_1, C_2)$
\begin{equation}\label{eq:IC}
    v'(\t, C_1, C_2)=-\V[X(\t)]\quad\mu-{\mbox{a.s.}}.
\end{equation}
In particular $v'(\cdot, C_1, C_2)\le 0$ $\mu$-a.s.. Equation~\eqref{eq:IC}
provides a valuable link between optimal contracts (from the point
of view of the agents) and the indirect utility functions.

In order to exploit the information contained in $\mu,$ firms must
choose catalogues that do not offer the agents incentives to lie
about their types. If firm $i$ intends to (partially) screen the market,
the subset $C_i(\t)\subset C_i$ of products from which it expects agents of type $\t$
to make their choices must satisfy
\be
    \E[X(\t)]-\t\V[X(\t)]-p(\t)\ge\E[Y]-\t\V[Y]-q,\quad\forall\,(X(\t), p(\t))\in C_i(\t),\, (Y,q)\in C_i.
\ee
Catalogues that satisfy this property are  called
{\sl incentive compatible}. Let
\be
    \mL(\t, C_i)\triangleq\textnormal{argsup}\big\{U(\t, X)-p\,\mid\, (X,p)\in C_i\big\},
\ee then $C_i$ is incentive compatible if and only if
$C_i(\t)=\mL(\t, C_i).$\footnote{It has been shown in~\cite{kn:rch}
that in general principal--agent models it is not possible to
perfectly screen the market. In most instances there is a
non--negligible set of agents who are pushed down to their
reservation utilities (bunching of the first type), and another one
where agents stay above their reservation levels, but they choose
the same product despite the fact of their different preferences
(bunching of the second type). In all likelihood, this behavior is
inherited by multi--firm models.} A catalogue where the products
intended for agents of type $\t$ yield at least their reservation
utility is said to be {\sl individually rational}. Since all
reservation utilities are zero, an individually rational catalogue
satisfies
\be
    \E[X(\t)]-\t\V[X(\t)]-p(\t)\ge 0\quad\forall\,(X(\t), p(\t))\in\mL(\t, C_i).
\ee
In the presence of the outside option, participation in the market is endogenously determined.
\subsection{Tie--breaking rules \& market segmentation}\label{subsec:TBRs_Market_Seg}
A crucial element  of multi--firm games is the presence of ties. There is no reason to assume
that for a given catalogue profile $(C_1, C_2)$ the sets $\mL(\t, C_1, C_2)$ ($\t\in\T$) will be singletons.
This fact renders the analysis considerably harder than it is for principal--agent games, where the Revelation Principle
allows for such assumption. There can be both inter-- and intra--firms ties, i.e. an agent may be indifferent between
two products that are offered by distinct firms, or maybe between two contracts that are offered by the same agency. In what follows
we study mechanisms via which ties are broken, as well as the partitions of $\T$ that are generated by tie--breaking.
In order to study the ties that originate from a catalogue profile $(C_1, C_2),$ we define
\be
    v(\t, C_i)=v_i(\t)\triangleq\sup\big\{U(\t, X)-p\,\mid\, (X,p)\in C_i\big\},
\ee
which are the one--catalogue analogues to $v(\t, C_1,
C_2).$ The $v_i$'s are related to $v(\t, C_1, C_2)$ via
\be
    v(\t, C_1, C_2)=\max\big\{v(\t, C_1),\,v(\t, C_2)\big\}.
\ee
Some important properties of the functions $v(\t, C_i)$ are summarized in Lemma~\ref{lm:convex_ind_ut} below. In a nutshell it states that the indirect utility functions generated by incentive compatible catalogues are convex, and that there is a crucial link between the derivatives of such functions at a given type and the variance of the contracts that such type may choose.

\begin{lemma}\label{lm:convex_ind_ut} Let $(C_1, C_2)$ be a catalogue profile and assume $\mL(\t, C_i)\neq\emptyset$ for all $\t\in\T,$ then:

\begin{enumerate}

\item The functions $v(\cdot, C_i):\T\to\re$ are convex.

\item If $C_i$ is incentive compatible then $-\V[X_i(\t)]\in\partial v(\t, C_i)$ for all $(X_i(\t), p_i(\t))\in C_i(\t).$ In particular, the equation $-\V[X_i(\t)]=v'(\t, C_i)$ holds $\mu$-a.s..

\end{enumerate}
\end{lemma}
\begin{Proof} The mapping $\t\mapsto U(\t, X)-p$ is affine on the $\t$-th coordinate, so $v(\t, C_i)$ is defined as the pointwise supremum of affine mappings and it is therefore convex (see for example Proposition 3.1 in~\cite{kn:et}). By assumption $\mL(\t, C_i)\neq\emptyset,$ an applying the Envelope Theorem as above we have that for any $(X_i(\t),
p_i(\t))\in\mL(\t, C_i),$ $-\V[X_i(\t)]=\in\partial v(\t, C_i).$
The assumption of incentive compatibility implies that if $(X_i(\t), p_i(\t))\in C_i(\t),$ then $-\V[X_i(\t)]=v'(\t, C_i)$ $\mu$-a.s..

\end{Proof}

If for $\t_0\in\T$ we have $v_i(\t_0)>v_{-i}(\t_0),$ then the
agents of type $\t_0$ will contract with firm $i.$ The set of
types that are indifferent between the firms' offers is \be
    \T_0\triangleq\big\{\t\in\T\,\mid\, v(\t, C_1)=v(\t, C_2)\big\}.
\ee
To avoid ambiguities we assume that if $v_i(\t)=v_{-i}(\t)=0$ then the
corresponding agents opt for the outside option. The market is then segmented in the sets
\be
\T_1\triangleq\{v_1>v_2\},\quad\T_2\triangleq\{v_2>v_1\}\quad{\mbox{and}}\quad\T_0.
\ee
In order to deal with
types in $\T_0$ whose indirect utility is not zero, we define
the set of tie--breaking rules as
\be
    F\triangleq\big\{f\in\Lp^0[a, 1]\,\mid\, 0\le f\le 1\big\}.
\ee
 From this point on, given a TBR $f$ we write $f_1=f$
and $f_2=1-f.$ Then $f_i\equiv 1$ on $\T_i,$ $f_i\equiv 0$ on $\T_{-i},$ and for $\t_0\in\T_0$  the proportion $f_i(\t_0)$
contracts with firm $i.$

In the sections below, we rephrase in as much as possible the
interaction of the firms in terms
of the indirect utilities generated by the catalogues they offer.
This provides a clearer understanding of the market's
segmentation, and in mathematical terms it allows us to use well
established convex analysis machinery. To this end we define \be
    \C_i\triangleq\Big\{v:\T\to\re_+\,\mid\, v\ge 0, \exists
    \{X(\t)\}_{\t\in\T}\subset\X_i\,\,{\mbox{s.t.}}\,\,
    v'(\t)=-\V[X(\t)]\Big\}.
\ee
These are the sets of all possible (single--firm) indirect utilities that can be generated from incentive compatible
catalogues contained in $\X_i.$ The incentive compatibility is reflected in the requirement $v'(\t)=-\V[X(\t)],$ as in Lemma~\ref{lm:convex_ind_ut}. We show below that these sets
exhibit convenient compactness properties.

\begin{proposition}\label{prop:uniform_bound_v}
The sets $\C_i$ are compact for the topology of uniform
convergence.
\end{proposition}
\begin{Proof} When firm $i$ designs a product line, it takes into account that
\be
    -M^2\le \E[X]-\t\V[X]\le M.
\ee
This implies that all prices $p\in P$ must be below $M$ to satisfy the
individual rationality constraint. In counterpart $-M^2\le p,$
otherwise all agents would be guaranteed an indirect utility above
their reservation utility. Since $\|X\|_2\le M$ for all $X\in\X_i,$
then $|v_i'|\le M$ for all $v_i\in\C_i,$ thus
\be
    0\le v_i\le 3\cdot\max\big\{M, M^2\big\}.
\ee
The convexity of the elements of the (closed) set $\C_i$ implies they are locally Lipschitz (see
for example~\cite{kn:ph1}); moreover, the $\Lp^2$--boundedness of $\X_i$ together with Equation~\eqref{eq:IC} imply the
Lipschitz coefficients are uniformly bounded. This in turn means that $\C_i$ is a bounded, closed and
uniformly equicontinuous family, which by the Arzel\`{a}--Ascoli is
then compact for the topology of uniform convergence.

\end{Proof}

\subsection{The social planer}\label{subsec:Social_Planer}

The fundamental theorems of welfare economics establish the equivalence between
competitive equilibria (in complete markets) and efficiency, in
the sense that frictionless competition leads to Pareto optimal
allocations of resources and viceversa. In contrast, in OTC markets
participants do not respond to given prices and therefore
optimality is an inadequate notion to study. Instead, in the
sections below we deal with the existence of {\sl Nash equilibria} and {\sl socially
efficient allocations}.

Whereas in perfectly competitive settings market forces interact as to eventually reach efficient outcomes, OTC markets may require the influence of a {\sl social planer} (a regulator) in order to achieve efficiency. This indirect market participant plays two important roles: First, he may choose to enforce certain kinds of TBRs in order to guarantee Nash and/or socially efficient outcomes. Second, he must make sure that individual rationality at the level of firms is preserved. This crucial implementability condition is a non--issue in competitive markets,
where equilibrium allocations are also individually rational.
Alternatively, given that socially efficient allocations are aggregately individually rational, the regulator could focus on efficiency and then establish a payment scheme among firms. These payment scheme would play the
role that equilibrium prices do in competitive markets.

In what follows we assume that we work on regulated OTC markets,
where the roles of the social planner are to seek that the market
settles for socially (aggregate) optimal allocations, and that the latter are individually rational for the agencies.

\section{Profit Maximization}\label{sec:ProfitMax}

In this section we analyze a non--cooperative game played among
profit--maximizing firms under adverse selection. Our model is an extension
into a multi--firm setting to the one studied in~\cite{kn:cet}. We show that if
{\sl efficient tie--breaking rules} are implemented (possibly
through the influence of a regulator), then the game possesses a
Nash equilibrium in mixed strategies.

\subsection{The firms' strategy sets \& payoff functions}\label{subsec:ProfitMax}

The theory of equilibria in multi--firm games developed (among others) in~\cite{kn:MaSt}, \cite{kn:Pa} and \cite{kn:PaMo2} requires the strategy sets to be compact metric spaces. Moreover, the agents' preferences must be continuous with respect to a topology that makes the strategy sets compact. In our view this implies that in general the strategy sets are actually closed and bounded sets of a finite--dimensional vector space. Following along the same lines, and with the aim of describing a profit--maximization model of OTC trading of derivatives contracts, we suppose that
firms structure the latter from a set of ``basic'' products. These products
are available in markets to which agents do not have access, and could be, for instance, OTC markets between ``high--rollers''.
The firms are then exposed to a cost when delivering each of these products.

We assume $\X_i$ is the closed convex hull of a finite number of
basic products $\{X_i^1,\ldots, X_i^{m_i}\},$ and that the cost to firm $i$ of delivering $X_i\in\X_i$ is given by a lower semicontinuous function
\be
    K_i:\X_i\to\re.
\ee
The per--contract profit of firm $i$ when it sells claim $X_i$
given the price schedule $p_i:\X_i\to\re$ is
\be
    p_i(X_i)-K_i(X_i).
\ee
Analyzing a multi--firm game played over non--linear price schedules would be daunting at best. Here the Revelation Principle (see for example
\cite{kn:PaMo2}) comes to our rescue. In a nutshell, it states that the game played over product--price catalogues,
i.e. elements of
\be
    \mP_i\triangleq\big\{C_i\subset\X_i\times P\,\mid\, C_i\,\,{\mbox{closed}}\big\},
\ee is as general as the one played over closed subsets of $\X_i$
and non--linear price schedules. We recall that $P\subset\re$ is the
compact set where feasible prices lie. We assume that firms compete
for the agents' business by offering product--price catalogues. In
other words, the strategy sets $\mP_i$ are the sets of all compact
subsets of $\X_i\times P.$ We endow $\mP_i$ with the Hausdorff metric $h.$ Since $\X_i\times
P$ is a compact metric space, so is $(\mP_i, h)$ (see, for
example~\cite{kn:ab}, Section 3.17); furthermore, Tychonoff's
theorem guarantees that $\mP\triangleq\mP_1\times\mP_2$ with the
corresponding product metric $h_p$ is also compact and metric. We write
$\B(\mP)$ for the Borel $\sigma$-algebra in $\mP.$ The following proposition allows us
to substitute, without loss of generality, a catalogue profile
$(C_1, C_2)$ for the agents' optimal choices associated with the
market situation $(C_1, C_2)$, i.e., for \be
    \Big(\bigcup_{\t}\mL(\t, C_1), \bigcup_{\t}\mL(\t, C_2)\Big)
\ee as these sets are closed and hence compact.

\begin{proposition}\label{prop:closed_union} For any catalogue $C_i\in\mP_i,$ the set $\bigcup_{\t}\mL(\t,
C_i)$ is closed.
\end{proposition}
\begin{Proof} Consider $(\bar{X}, \bar{p})\in{\mbox{cl}}(\bigcup_{\t}\mL(\t,
C_i)),$ then there exists a sequence $\{(X_n, p_n)\}\subset \bigcup_{\t}\mL(\t,
C_i)$ such that
\be
    (X_n, p_n)\xrightarrow{\|\cdot\|_2\times |\cdot|} (\bar{X} ,\bar{p}).
\ee
By construction $(X_n, p_n)\in\mL(\t, C_i)$ for some $\t\in\T,$ call it $\t_n.$ Passing to a subsequence if necessary we may assume there is $\bar{\t}\in\T$ such that $\t_n\to\bar{\t}$ and $(X_n, p_n)\to (\bar{X}, \bar{p})$ pointwise. If $(\bar{X}, \bar{p})\in \bigcup_{\t}\mL(\t, C_i)$ we are done, so let us assume the contrary, thus
\begin{equation}\label{eq:not-in-L}
    \E[\bar{X}]-\bar{\t}\V[\bar{X}]-\bar{p}< \E[X]-\bar{\t}\V[X]-p\quad\forall\,(X, p)\in\mL(\bar{\t}, C_i).
\end{equation}
By definition
\be
    \E[X_n]-\t_n\V[X_n]-p_n\ge \E[X]-\t_n\V[X]-p\quad\forall\,(X, p)\in C_i.
\ee
However, the (a.s.) convergence of the sequence $\{(X_n, p_n)\}$ implies
\be
    \E[\bar{X}]-\bar{\t}\V[\bar{X}]-\bar{p}\ge \E[X]-\bar{\t}\V[X]-p\quad\forall\,(X, p)\in C_i,
\ee
which would imply $(X, p)\in\mL(\bar{\t}, C_i),$ contradicting Equation~\eqref{eq:not-in-L}.

\end{Proof}

\noindent The maximal attainable profit for firm $i$ from type $\t$ given
the (incentive compatible) catalogue profile $(C_1, C_2)$ is
\be
    \pi_i(\t, C_1, C_2)\triangleq\max\big\{(p-K(X))\i_{\{v_i\ge
    v_{-i}\}}(\t)\,\mid\, (X, p)\in C_i\big\}.
\ee Here the functions $v_i$ are as defined in
Section~\ref{subsec:TBRs_Market_Seg}. The functions $\pi_i$
are the building blocks of the firms' payoff functions and enjoy the
following important continuity property:

\begin{proposition}\label{prop:lsc}(Page Jr.~\cite{kn:Pa}) The maps
$\pi_i:\T\times\mP\to\re$ are upper semicontinuous on $\mP$ and
$\B(\T)\times\B(\mP)$-measurable.
\end{proposition}
For a given TBR $f,$ the payoff function of firm $i$ is given by
\be
    \Pi_i(C_1, C_2)\triangleq\int_{\T}\pi_i(\t, C_1, C_2)f_i(\t)\mu(d\t).
\ee
These payoff functions exhibit a highly discontinuous behavior, due to the presence of ties. However,
the effect of each type--wise trade on the overall income of the firm is linear and easily
quantifiable. Moreover, only agents who make a positive contribution to a firm's revenues contract with it. This last
fact plays a role in the analysis of {\sl payoff security} contained in Section~\ref{subsubsec:UPSec}.

\subsection{Efficient tie--breaking rules \& reciprocal upper semicontinuity}\label{subsec:ETR_RUSC}

Roughly speaking, a TBR is said to be {\sl efficient} if an agent
contracts with a firm if and only if the latter values her as a
customer at least as much as the competition. As we show below,
this is equivalent to saying that an efficient TBR maximizes the
aggregate profit for a given market situation $(C_1, C_2).$

\begin{definition} Given a catalogue profile $(C_1, C_2),$ a tie--breaking
rule $f\in F$ is called {\bf{efficient}} if

\begin{itemize}

\item $\pi_i(\t, C_1, C_2)\ge \pi_{-i}(\t, C_1, C_2)\quad{\mbox{implies}}\quad f_i(\t)>0,$

\item $\pi_i(\t, C_1, C_2)< \pi_{-i}(\t, C_1, C_2)\quad{\mbox{implies}}\quad f_i(\t)=0.$

\end{itemize}
\end{definition}
The definition above is equivalent to saying that $f^*$ is
efficient given the catalogue profile $(C_1, C_2)$  if and only if
\be
    \sum_i\Pi_i(f^*, C_1, C_2)=\sup_{f\in F}\sum_i\Pi_i(f, C_1, C_2),
\ee
which shows that in general efficient TBRs are endogenously determined. In mathematical terms, efficient TBRs
yield certain upper semicontinuity properties to the payoff
functions of the firms, which are necessary to guarantee the
existence of Nash equilibria.

\begin{definition} A game $\big\{(\Pi_i, \mP_i)\big\}$ is said to be {\bf{
reciprocal upper semicontinuous}} (RUSC) for a given TBR $f$ if the mapping
\be
(C_1, C_2)\mapsto\sum_i \Pi_i(f, C_1, C_2)
\ee
is upper semicontinuous.
\end{definition}
The notion of RUSC games was introduced by Dasgupta and Maskin
in~\cite{kn:DM} (labeled as complementary discontinuous or
u.s.c--sum games) and later generalized by Reny in~\cite{kn:Re} in
order to prove the existence of Nash equilibria in certain
discontinuous games. Intuitively, in a RUSC game firm $i$ can only
approximate the payoff corresponding to a market situation $(C_1,
C_2)$ by actually playing $C_i.$ We should note that Reny's
definition is slightly stronger, but the one above
(from~\cite{kn:DM}) is sufficient for our needs. Furthermore,
Proposition 5.1 in~\cite{kn:Re} tells us that RUSC is inherited by
the mixed extension of a game.

\begin{lemma}\label{lm:RUSC}(Page Jr. \& Monteiro~\cite{kn:PaMo3}) If the game $\big\{(\mP_i, \Pi_i)\big\}$ is played using
efficient TBRs, then it is RUSC.
\end{lemma}
\begin{Proof} Consider the mapping $(C_1, C_2)\mapsto\sum_i \Pi_i(f^*, C_1, C_2).$ If $f^*$ is efficient then we have
\begin{eqnarray*}
\sum_i \Pi_i(f^*, C_1, C_2) & = & \sup_{f\in F}\sum_i\Pi_i(f, C_1, C_2)\\
                            & = & \sup_{f\in F}\sum_i\int_{\T}\pi_i(\t, C_1, C_2)f_i(\t)\mu(d\t)\\
                            & = & \int_{\T} \max_i \pi_i(\t, C_1, C_2)\sum_i f_i(\t)\mu(d\t)\quad{\mbox{for any}}\quad f\in F\\
                            & = & \int_{\T} \max_i \pi_i(\t, C_1, C_2)\mu(d\t).\\
\end{eqnarray*}
It follows from Proposition~\ref{prop:lsc},  that $(C_1, C_2)\mapsto\max_i \pi_i(\t, C_1, C_2)$ is upper semicontinuous, hence so is $\sum_i \Pi_i(f^*, C_1, C_2).$

\end{Proof}

\noindent Notice that the definition of efficient TBR, together with the linear aggregation of type-wise profits, implies that the mapping $(C_1, C_2)\mapsto\sum_i \Pi_i(f, C_1, C_2)$ is independent of the TBR played if the latter it is efficient.
\subsection{Existence of Nash equilibria}
In this section we study necessary conditions for the existence of
Nash equilibria in the agencies' game. We cannot employ the
canonical results of Debreu, Glicksberg and Fan, given the
discontinuities introduced by the TBRs. Instead we rely on the
results of Page Jr. \& Monteiro~\cite{kn:PaMo2} and
Reny~\cite{kn:Re}. Since the payoff functions $\Pi_i$ are not
quasiconcave, one cannot prove the existence of pure--strategies
equilibria using the state--of--the--art results available in the
literature. Instead we analyze the existence of mixed--strategies
equilibria. In order to do so we show {\sl uniform payoff security}
of the game. This notion allows one to conveniently test the {\sl
payoff security} of a game's mixed extension.

\subsubsection{Defining the mixed--strategies game}

To define a {\sl mixed catalogue game}, we let $\mm(\mP_j)$ be the set of probability measures on $(\mP_i, \B(\mP_i)).$  We endow the {\sl mixed catalogue strategy sets} $\mm(\mP_i)$ with the $\sigma(X^*, X)$-topology generated by the dual pair $\langle \Lp^0(\mP_i, \B(\mP_i)), \mm(\mP_i)\rangle.$ The  game is played as follows: Each agency chooses an element $\lambda_i\in\mm(\mP_i),$ and its profit under the profile $\lambda=(\lambda_1, \lambda_2)$ is given by
\begin{equation}\label{eq:mixed-profit}
    \P_i(\lambda)\triangleq\int_{\mP}\Pi_i(C_1, C_2)\lambda(dC),
\end{equation}
where $\lambda(dC)=\lambda_1(dC_1)\lambda_2(dC_2)$ is the corresponding product measure. By Proposition~\ref{prop:lsc}  the integrand is measurable, thus expression~\eqref{eq:mixed-profit} is well defined. The game $\big\{(\mm(\mP_i), \P_i)\big\}$ is the {\sl mixed catalogue game} that extends $\big\{(\mP_i, \Pi_i)\big\}.$

\begin{definition}\label{def:NashEquilMixed}
A mixed profile $\lambda$ is a {\bf Nash equilibrium} for the game $\big\{(\mm(\mP_i), \P_i)\big\}$ if
\be
\P_i(\lambda_i,\lambda_{-i})\ge \P_i(\lambda_i', \lambda_{-i})\quad{\textnormal{ for all}}\quad\lambda_i'\in\mm(\mP_i).
\ee
\end{definition}
A Nash equilibrium in mixed strategies for $\big\{(\mm(\mP_j), \P_i)\big\}$ can be viewed as an equilibrium in pure strategies for its mixed extension.

\subsubsection{Uniform payoff security}\label{subsubsec:UPSec}

A game in which the players' strategy sets are compact subsets of a
topological space and their payoff functions are bounded is called
{\sl compact}. The mixed extension of the game in hand is a compact
game. This is due to the fact that  the sets $\mm(\mP_i)$ are
closed, convex and bounded subsets of $(\Lp^0(\mP_i, \B(\mP_i)))^*,$
thus by the Banach--Alaoglu theorem they are
$\sigma(X^*,X)$-compact. Reny provides sufficient conditions for
existence of a Nash equilibrium in a compact game in the presence of
discontinuous payoff functions. A key requirement is that the game
is {\sl payoff secure}. Intuitively this means that small deviations
on the part of firm $i$'s competitors can be answered to with a
deviation that leaves firm $i$ within a small interval of its profit
prior to deviations.

\begin{definition}\label{def:payoffSec} Given the payoffs $\Pi_i:\mP_i\to\re,$
the catalogue game $\big\{(\mP_i, \Pi_i)\big\}$ is {\bf{payoff secure}} if for all $(C_1, C_2)\in\mP,$  and $\epsilon>0$ there exist $C_i^*\in\mP_i$ and $\delta>0$  such that
\begin{equation*}
\Pi_i(C^*_i, C_{-i}')\ge \Pi_i(C_i, C_{-i})-\epsilon
\end{equation*}
for all $C_{-i}'\in \big\{D\in\mP_{-i}\,\mid\, h_p(D, C_{-i})<\delta\big\}.$
\end{definition}
Payoff security of a game does not imply the same property for its mixed extension; however, such is the case with {\sl uniform payoff security} (see Theorem 1 in~\cite{kn:PaMo3}). Proving uniform payoff security of a game is simpler than dealing with the weak$^*$-topology to show payoff security of its mixed extension.

\begin{definition}\label{def:UnifpayoffSec}
The catalogue game $\big\{(\mP_i, \Pi_i)\big\}$ is {\bf{uniformly payoff secure}} if for all $i= 1, 2,$ $C_i\in\mP_i$ and $\epsilon>0$ there exist $C_i^*$ such that for all $C_{-i}\in\mP_{-i}$ there exists $\delta>0$ that satisfies
\be
    \Pi_i(C^*_i, C_{-i}')\ge \Pi_i(C_i, C_{-i})-\epsilon
\ee
for all $C_{-i}'\in\big\{D\in\mP_{-i}\,\mid\, h_p(D, C_{-i})<\delta\big\}.$
\end{definition}

\begin{proposition}\label{pr:UPSec}
The game $\big\{(\mP_i, \Pi_i)\big\}$ is uniformly payoff secure (hence the game $\big\{(\mm(\mP_j), \P_i)\big\}$ is payoff secure).
\end{proposition}
\begin{Proof} Let $C_i\in\mP_i$ and $\epsilon>0$ and define
\be
C_i^{\epsilon}\triangleq\big\{(X, p-\epsilon)\,\mid\, (X, p)\in C_i\quad{\mbox{and}}\quad p-\epsilon\ge K_i(X)\big\}\cup\big\{(0, 0)\big\},
\ee
i.e. $C_i^{\epsilon}$ is obtained from $C_i$ by keeping all contracts that allow for an $\epsilon$--decrease in prices without this rendering a negative price--cost benefit. All other contracts are replaced by $\big\{(0, 0)\big\}.$ Lemma~\ref{lemma:equicont} implies there is $\delta>0$ such that if $d\big((X, p), (X', p')\big)<\delta$ (where $d(\cdot, \cdot)$ is the distance generated by $\|\cdot\|_2\times|\cdot|$) then for all $\t\in\T$
\begin{equation}\label{eq:hhhhh}
|U(\t, X)-p-U(\t, X')+p'|<\epsilon.
\end{equation}
Assume first that $\pi_i(\t, C_i, C_{-i})>\epsilon$ and let $(X, p)\in C_i$ be such that $U(\t, X)-p=\pi_i(\t, C_i, C_{-i}).$ Then $(X, p-\epsilon)\in C_i^{\epsilon}.$ Now consider $C_{-i}'$ such that $h_p(C_{-i}, C_{-i}')<\delta,$ and let $(X', p')\in C_{-i}'.$  By definition there exist $(Y, q)\in C_{-i}$ such that  $d\big((Y, q), (X', p')\big)<\delta,$ and by Equation~\eqref{eq:hhhhh}
\be
U(\t, X)-(p-\epsilon)\ge U(\t, Y)-(q-\epsilon)>U(\t, X')-p'.
\ee
The latter implies that $(X', p')\notin\mL(\t, C_i^{\epsilon}, C_{-i}')$ and in turn that an efficient TBR $f$ for $(C_i^{\epsilon}, C_{-i}')$ satisfies $f_i(\t)=1.$ Thus
\be
\pi_i(\t, C_i^{\epsilon}, C_{-i}')f_i(t)\ge \pi_i(\t, C_i^{\epsilon}, C_{-i}').
\ee
The inequality above is trivially fulfilled if $\pi_i(\t, C_i, C_{-i})\le\epsilon.$ We have that for any deviation $C_{-i}'$ such that $h_p(C_{-i}, C_{-i}')<\delta$ and for any efficient TBR $f$
\begin{eqnarray*}
\Pi_i(C^{\epsilon}_i, C_{-i}') & \ge & \int_{\T} \pi_i(\t, C_i^{\epsilon}, C_{-i}')f_i(\t)\mu(d\t)-\epsilon\\
                               &  =  & \Pi_i(C_i, C_{-i})-\epsilon.\\
\end{eqnarray*}

\end{Proof}

The proof of Proposition~\ref{pr:UPSec} is a particular case of the proof of Theorem 1 in~\cite{kn:PM4}. We have included it here because it showcases how the linear aggregation of type--wise profits is key to the existence of Nash equilibria. The simple yet far reaching construction of $C_i^{\epsilon}$ is only possible in such case. Our result on existence of Nash equilibria relies on the following

\begin{thm}(Reny~\cite{kn:Re})\label{theo:Reny} Suppose that $G =\{(X_i, u_i)\}_{i=1}^m$
is a compact game which is also quasiconcave, reciprocally upper semicontinuous and payoff secure, then it possesses a pure--strategies Nash equilibrium.
\end{thm}
It follows from Lemma \ref{lm:RUSC} that the (compact) game $\big\{(\mm(\mP_i), \P_i)\big\}$ is RUSC. The payoff functions are linear in each firm's strategy, hence quasiconcave. Proposition~\ref{pr:UPSec} then allows us to apply Theorem~\ref{theo:Reny} to the mixed extension of the game  $\big\{(\mP_i, \Pi_i)\big\}$ to obtain the following
\begin{thm}\label{thm:main-profit}
The game $\big\{(\mP_i, \Pi_i)\big\}$ possesses a mixed--strategies Nash
equilibrium.
\end{thm}

\section{Risk Minimization}\label{sec:RiskMin}

In this section we analyze the risk minimization problem faced by
firms that have some initial risky endowments. The goal of
the firms is to lay off parts of their risk on individual agents.
This is done via OTC trading of derivatives contracts. The model presented here
is the multi--firm version of the one introduced in~\cite{kn:hm}.
In contrast with the previous section, we do not seek to prove the
existence of Nash equilibria, but rather of {\sl socially
efficient allocations}. The main difficulty towards guaranteeing Nash equilibria stems
from the non--linear, per--type impact on the firms' risk assessments. In the previous
sections, the fact that any agent contracting with an agency had a positive
impact on its revenues was essential to show uniform payoff security and RUSC.
This is no longer the case for risk--minimizing firms. Here the risk associated to the aggregate
of a firm's initial position  plus a contract  could be lower than the firm's initial
risk assessment, but once all positions are taken into account, the firm might be better off not engaging
the agents who would have chosen such contract.
\subsection{The firms' strategy sets \& risk assessments}\label{sec:Firms}

The initial risky endowment of each firm is represented by a
random variable $W_i\in\Lp^2\left(\Omega, {\cal{F}}, \Prob
\right)$. Firm $i$ assesses its risk exposure using a convex and
law invariant risk measure
\be
    \varrho_i:\Lp^2\left(\Omega, {\cal{F}}, \Prob\right)\to\re\cup\{+\infty\},
\ee
which  has the Fatou property. We refer the reader to Appendix
~\ref{app:RM} for a discussion on these maps, as well as related
references. We restrict the firms' choices to catalogues of the
form
\be
    C_i=\big\{(X_i(\t), p_i(\t))\big\}_{\t\in\T}.
\ee
These are not direct revelation mechanisms: even if firm $1$ were to offer the
individually rational and incentive compatible catalogue $C_1,$
the presence of firm $2$ might dissuade agents of type $\t_0$ from
choosing $(X_1(\t_0), p_1(\t_0)).$
For a given catalogue profile $(C_1, C_2)$ and a given TBR $f,$
the position of firm $i$ after trading is
\be
    W_i-\int_{\T_i}(p_i(\t)-X_i(\t))\mu(d\t)-\int_{\T_0}(p_i(\t)-X_i(\t))f_i(\t)\mu(d\t),
\ee
where $\T_i$ and $\T_0$ are as defined in Section~\ref{subsec:TBRs_Market_Seg}.
If we write $p_i(\t)=\t v_i'(\t, C_1, C_2)-v_i(\t, C_1,
C_2),$ and we rename $X(\t)=X(\t)-\E[X(\t)],$\footnote{This is possible due to the translation invariance of $\varrho_i,$ and fully characterizes $p_i$ in terms of $v_i.$} then firm $i$'s
risk assessment after trading is
\be
    A_i( C_1, C_2, f)\triangleq R_i(C_1, C_2, f)-I_i((C_1, C_2, f),
\ee
where
\be
    R_i(C_1, C_2, f)\triangleq\varrho_i\bigg(W_i-\int_{\T_i}X_i(\t)\mu(d\t)
    -\int_{\T_0}X_i(\t)f_i(\t)\mu(d\t)\bigg),
\ee
denotes firm $i$'s risk and the associated income is given by
\be
    I_i((C_1, C_2, f)\triangleq\int_{\T_i}(\t v_i'(\t)-v_i(\t) \mu(d\t)
    +\int_{\T_0}(\t v'_i(\t)-v_i(\t))f_i(\t)\mu(d\t).
\ee
As we have seen before, when a catalogue profile $(C_1, C_2)$ is presented to the agents, the corresponding indirect utility functions $(v_1, v_2)$ show how the market is segmented. The more interesting set is $\T_0,$ i.e. the set of indifferent agents. Within this set, there are two intrinsically different situations. First, $v_1$ and $v_2$ may be identical over a non--negligible subset. These could be regarded as ``true'' ties, in the sense that they will have different impacts on the firms' risk assessments under different TBRs. Second, there could be types for which $v_1$ and $v_2$ are secant. We show below that given our assumptions on $\mu,$ these types do not have a direct impact on the aggregate incomes (in fact they do not impact the $A_i$'s, but we do not require this for our arguments). However, they do indicate the points where agents switch from contracting with one firm to the other one, in other words they show where the customers' preferences shift.

\begin{definition} Given a pricing schedule $(v_1, v_2)\in\C_1\times\C_2,$ we say that there is a {\bf shift in the customers'
preferences} at $\t\in{\mbox{int}}(\T)$ if
\be
    (v_1(\t_{-})-v_2(\t_{-}))\cdot(v_1(\t_{+})-v_2(\t_{+}))<0.
\ee
\end{definition}

\begin{proposition}\label{prop:pref_shift} For any pricing schedule $(v_1, v_2)\in\C_1\times\C_2$ the set $\T_s$ of shifts in the customers'
preferences satisfies:

\begin{enumerate}

\item $\T_s$ has at most countably many elements.

\item The derived set of $\T_s$ has measure zero.

\end{enumerate}
\end{proposition}
\begin{Proof}  The set $\T_s(v_1, v_2)$ is the union of the closures of the sets
\be
    \T_s^1\triangleq\left\{\t\in\T\,\mid\,v_1(x)=v_2(x), v_1'(x)\neq v_2'(x)\right\}
\ee
and
\be
    \T_s^{\epsilon}\triangleq\left\{\t\in\T\,\mid\,v_1(x)=v_2(x),\, \,v_1'(x)=v_2'(x),\,
    |v_1(\tilde{x})-v_2(\tilde{x})|>0\,\,\forall\,\tilde{x}\in B(\epsilon, x)\setminus\{x\},\,\epsilon>0\right\}.
\ee
The set $\T_s^1$ is denumerable and nowhere dense. This follows directly from the convexity of $v_1$ and $v_2$ and the fact that the corresponding  supporting planes to graph$\{v_1\}$ and graph$\{v_2\}$ at $(\t, v_1(\t))$ are secant. By definition $\T_2$ is denumerable and nowhere dense, since any two of its elements can be separated. Therefore cl$\{\T_s^1\}$ and cl$\{\T_s^{\epsilon}\}$ are themselves denumerable and nowhere dense.

\end{Proof}

As a result of the preceding proposition we see that the set of pre--images
of the crossings of the graphs of two functions $v_1\in\C_i$ and $v_2\in\C_2$ has $\mu$--measure
zero, which yields the following

\begin{corollary}\label{cor:derivatives_equal}
Let $v_1\in\C_1$ and  $v_2\in\C_2,$ then the set $\{v_1=v_2\}\cap\{v_1'\neq v_2'\}$ is of $\mu$--measure zero.
\end{corollary}
As a consequence of Corollary~\ref{cor:derivatives_equal}, we have
that the aggregate income of the firms is independent of the TBR,
namely
\be
    \sum_i I_i(C_1, C_2, f)=\int_{\T_0}(\t v(\t)-v'(\t))\mu(d\t)
    +\sum_i\int_{\T_i}(\t v_i'(\t)-v_i(\t) \mu(d\t).
\ee
Finally we define
\be
    \A(C_1, C_2, f)\triangleq\sum_i A_i( C_1, C_2, f).
\ee

\subsection{Socially efficient allocations}
A market situation $(C_1^*, C_2^*),$ together with a TBR $f^*$ is
said to be a {\sl socially efficient allocation}\footnote{From
this point on we use the shorthand SEA to refer to a socially
efficient allocation.} if it minimizes the aggregate risk in the
economy and if it is individually rational at the agencies' level; in other words if
$\A(C_1^*, C_2^*, f^*)$ solves the problem
\be
    \Psi\triangleq\inf_v\inf_f\inf_X\big\{\A(C_1, C_2, f)\,\mid\,
    -\V[X_i(\t)]=v_i'(\t),\, A_i(C_1, C_2, f)\le\varrho_i(W_i)\big\}.
\ee
As it was the case in Lemma~\ref{lm:convex_ind_ut}, the variance constraints $-\V[X_i(\t)]=v_i'(\t)$  capture
the incentive compatibility of the catalogues $C_i,$ while $A_i(C_1, C_2, f)\le\varrho_i(W_i)$ is  firm $i$'s
individual rationality constraint. As we mentioned in Section~\ref{subsec:Social_Planer}, in
absence of a competitive market one cannot rely on equilibrium
pricing to take care of the issue of individual rationality. It is
therefore necessary to work under the constraints $A_i(C_1, C_2,
f)\le\varrho_i(W_i),$ for otherwise one could end up with
allocations that are optimal on the aggregate level, yet
unenforceable. An alternative would be to establish a
cash--transfer system, which should be supervised by the
regulator. We comment further on the latter in
Section~\ref{subsubsec:Ind_Rat_revisited}. In the remainder of
this section we study the existence of SEAs, and we show that the
presence a regulator is required in order for such allocations to
be attainable and/or implementable. Our existence result depends
heavily on the implementation of efficient TBRs.

\begin{definition}
Let $C=(C_1, C_2)$ be a catalogue profile. A tie--breaking rule
$\bar{f}\in F$ is {\bf{efficient}} for $C$ if
\be
    \A(C_1, C_2, \bar{f})=\inf_{f\in F}\big\{\A(C_1, C_2, f)\big\}.
\ee
\end{definition}
From the definition above one observes that efficient TBRs are
endogenously determined. Here we encounter the first need for the
social planner in our risk--minimization setting, as it could be
the case that unless regulated, efficient TBRs would not be
implemented.
\subsubsection{Minimizing the risk for fixed incomes and tie--breaking rule}\label{section:Min_risk_fixed_incomes}

In a first step, we fix $(v_1, v_2)\in\C_1\times\C_2$ (hence the firms' incomes), as
well as $f\in F.$ We shall abuse notation slightly and write $R_i(v_1,
v_2, X_i, f)$ for $R_i(C_1, C_2, f).$ We then analyze the problem
\begin{small}
\be
    \Psi_1\triangleq\inf_{(X_1, X_2)}\Big\{\sum _i R_i(v_1, v_2, X_i, f)\,\big|\,
    -\V[X_i(\t)]=v_i'(\t)\,\,{\textnormal{and}}\,\,A_i(v_1, v_2, X_i, f)\le\varrho_i(W_i) \Big\}.
\ee
\end{small}
This problem can be decoupled into the sum of the infima,
since the choice of $X_i$ bears no weight on the evaluation of
$R_{-i}.$ Hence we must study the solution(s) to the
following single--firm problems:
\be
    \inf_{X_i}\left\{R_i(v_i, v_{-i}, X_i, f)\,\big|\,-\V[X_i(\t)]=v_i'(\t)\,\,{\textnormal{and}}\,\,
    A_i(v_1, v_2, X_i, f)\le\varrho_i(W_i)\right\}.
\ee
To deal with the individual rationality constraints, we define
 \be
    \X_i^{v, f}\triangleq\big\{X_i\in\X_i\,\mid\, A_i(v_1, v_2, X_i,f)\le\varrho_i(W_i)\big\},
 \ee
which is the set of individually rational products for the given price schedule $(v_1, v_2),$ and
 \be
    \X_i^{k_i^v}\triangleq\big\{X\in\X_i\,\mid\,\|X_i\|_2^2\le k_i^v\big\}
 \ee
where $k_i^v\triangleq v_i(a)-v_i(0).$ The set $\X_i^{k_i^v}$ contains all the products that can be structured as to
construct incentive compatible catalogues given $(v_1, v_2),$ with $k_i^v$ providing a bound to the $\Lp^2$--norm of the products via the constraint $-\V[X_i(\t)]=v_i'(\t).$ We elaborate further into the $k_i^v$'s in Appendix~\ref{app:MP1}, where we also
provide an outline of the proof of existence of solutions to $\Psi_1$ stated in Lemma~\ref{lm:MP1}. The proof is analogous to that of Theorem 2.3 in~\cite{kn:hm}, and we include it for completeness.

\begin{lemma}\label{lm:MP1} For $v\in\C_1\times\C_2$ and $f\in F$ given, if $\X_i^{k_i^v}\cap\X_i^{v, f}\neq\emptyset$ then there exist $X_i^{v, f}\in\X_i$  such that
\be
    \Psi_1=R_1(v_1, v_2, X_1^{v, f}, f)+R_2(v_1, v_2, X_2^{v, f}, f).
\ee
\end{lemma}
\begin{remark}
It should be mentioned that individual rationality may be verified ex--post. If a solution $X_i^{v, f}$ to the minimization problem
\be
    \inf_{X_i}\left\{R_i(v_i, v_{-i}, X_i, f)\,\big|\,-\V[X_i(\t)]=v_i'(\t)\right\}
\ee
satisfies $A_i(v_1, v_2, X_i^{v, f}, f)\le\varrho_i(W_i),$ then $\X_i^{k_i^v}\cap\X_i^{v, f}\neq\emptyset.$ If on the contrary, $A_i(v_1, v_2, X_i^{v, f}\, f)>\varrho_i(W_i),$ then the solution set of $\Psi_1$ is empty.
\end{remark}

In order to establish the existence of an efficient TBR it is
important to characterize the optimal contracts $X_i^{v,
f}$. Specifically, we show below that $X_i^{v, f}$ can be
multiplicatively decomposed into a $\t$--dependent function and an
$\omega$--dependent random variable. To this end, we construct the
Lagrangian associated to minimizing $R_i(v_i, v_{-i}, X_i, f)$
subject to the moment conditions $\E [X_i(\t)]=0$ and
$\V[X_i(\t)]+v_i'(\t)=0,$ and we compute the Frech\'et differential
of $R_i$ at $X_i$ in the direction of $h\in\X_i:$
\begin{small}
\be
    R_i'(X_i)h =
    \varrho_i'\bigg(W_i-\int_{\T_i}X_i(\t)\mu(d\t)-\int_{\T_0}X_i(\t)f_i(\t)\mu(d\t)\bigg)
    \bigg(-\int_{\T_i}h(\t)d\t-\int_{\T_0}h(\t)f_i(\t)\mu(d\t)\bigg).
\ee
\end{small}Since, for all $H \in L^2(\O, \Prob),$ the map $K\mapsto\varrho_i'(H)K$ is linear, it follows from the Riesz
representation theorem that there is a random variable $Z_{X_i^{v,
f}} \in L^2(\O, \Prob)$ such that \be
    R_i'(X_i)h =\int_{\O}Z_{X_i^{v, f}}\Big(-\int_{\T_i}h(\t)d\t-\int_{\T_0}h(\t)f_i(\t)\mu(d\t)\Big)d\Prob.
\ee
Let $g:F\to\Lp^0(\T,\mu)$ be given by
\be
    g(f)=\i_{\T_i}+\i_{\T_0}f.
\ee
The operator $R_i$ has an extremum at $X_i$ under our
moment constraints if there exist Lagrange multipliers $\lambda_i,
\eta_i \in\Lp^2(\T,\mu)$ such that
\be
    \int_{\O}\int_{\T}h(\t)\left(-Z_{X_i^{v, f}}g(f_i)(\t)+\eta_i(\t)
    +2\lambda_i(\t)X_i(\t)\right)\mu(d\t) d\Prob=0.
\ee
for all $h\in\X_i$. Since $(\t,\o)\to
-Z_{X_i^{v, f}}g(f_i)(\t)+\eta_i(\t)+2\lambda_i(\t)X_i(\t)$ is an integrable function, the DuBois--Reymond lemma implies
\begin{equation}\label{eq:DuBois-Reymond}
    -Z_{X_i^{v, f}}g(f_i)(\t)+\eta_i(\t)+2\lambda_i(\t)X_i(\t)=0.
\end{equation}
Using the moment conditions  $\E_\Prob[X_i^{v, f}(\t)]=0$ and $\V[X_i^{v, f}(\t)]=-v_i'(\t)$ we obtain
\be
    \eta_i(\t)=\E_\Prob\big[Z_{X_i^{v, f}}\big]g(f_i)(\t)\quad\textnormal{and}\quad\lambda_i(\t)=\dfrac{g(f_i)(\t)\sqrt{\V\big[Z_{X_i^{v, f}}\big]}}{2\sqrt{-v_i'(\t)}}.
\ee
Inserting the expressions for the Lagrange multipliers into Equation~\eqref{eq:DuBois-Reymond} yields
\begin{equation}\label{eq:representation}
    X_i^{v,f}(\t)=\sqrt{-v_i'(\t)}\bar{Z}_i^{v, f},
\end{equation}
where
\be
    \bar{Z}_i^{v, f}\triangleq \Big(Z_{X_i^{v, f}}-\E\big[Z_{X_i^{v, f}}\big]\Big)\big/\sqrt{\V\big[Z_{X_i^{v, f}}\big]}.
\ee
Equation~\eqref{eq:representation} shows that the minimizers of problem $\Psi_1$ form collinear families in $\X.$ Moreover, the randomness stemming from $(\O, \Prob)$ and and the one induced by $(\T, \mu)$ are decoupled. This property will prove to be key in Section~\ref{subsubsection:ETR_Risk}, where we show the existence of efficient TBRs.
The previous discussion yields the following characterization result:
\begin{proposition}\label{prop:representation_optimal_contracts}
Let $({X}_1^{v, f}, {X}_2^{v, f})$ be a solution to
$\Psi_1.$ Then ${X}_i^{v, f}$ takes the form
\be
    {X}_i^{v, f}(\t,\o) = \sqrt{-v_i'(\t)}\bar{Z}_i^{v, f}(\o)
\ee
for some normalized random variable $\bar{Z}_i^{v, f}$ on $(\Omega, {\cal F}, \Prob)$.
\end{proposition}
If either ${X}_1^{v, f}$ or  ${X}_2^{v, f}$ do not satisfy $A_i(v_1, v_2, X_i^{v, f},
f)\le\varrho_i(W_i),$ then using the standard convention $\inf\{\emptyset\}=\infty$ we would have $\Psi_1=\infty.$
In terms of the program $\Psi,$ this guarantees that only pricing schedules and TBRs that offer the
possibility of constructing individually rational catalogues stand
a chance to be chosen.
\subsubsection{Existence of efficient tie--breaking rules}\label{subsubsection:ETR_Risk}

In this section we show that for a given price schedule $(v_1,
v_2)$ there is a TBR $f^v$ such that the corresponding optimal product lines $(X_1^{v,f^v},
X_2^{v,f^v})$  minimize the aggregate
risk evaluation of the firms.
For $v=(v_1, v_2)\in\C_1\times\C_2$
we define
\be
    \tilde{R}_i(v_1, v_2, f)
    \triangleq\varrho_i\bigg(W_i-Z_i^{v,f}(\omega)\,\Big(\int_{\T_i}
    \sqrt{-v_i'(\t)}d\t+\int_{\T_0}f_i(\t)\sqrt{-v_i'(\t)}\mu(d\t)\Big)\bigg)
\ee
and
\be
    F_v\triangleq\Big\{f\in F\,\mid\, \sum_i \tilde{R}_i(v_1, v_2, f)<\infty\Big\}.
\ee
We then have to solve the problem
\be
    \Psi_2\triangleq \inf_{F_v} \sum_i \tilde{R}_i(v_1, v_2, f).
\ee
If $F_v=\emptyset,$ then as above $\Psi_2=\infty.$ Otherwise we must verify
the lower semicontinuity of the mapping
\be
    f\mapsto\tilde{R}_1(v_1, v_2, f)+\tilde{R}_2(v_1, v_2, f),\quad f\in F_v.
\ee
To this end consider a minimizing sequence $\{f^n\}\subset F_v.$
The Banach--Alaoglu theorem guarantees that $\{f^n\}$ is  weakly
convergent\footnote{We omit writing ``up to a subsequence'' when exploiting the compactness of sets, but it
should of course be kept in mind.} to some $f^v\in F_v.$ Since
$V_i(\t)\triangleq\sqrt{-v_i'(\t)}$ belongs to $\Lp^2(\T, \mu)$ for all
$v_i\in\C_i,$ then
\begin{equation}\label{eq:conv_TBR}
    \int_{\T_0}f_i^n(\t)V_i(\t)d\t\to \int_{\T_0}{f_i^v}(\t)V_i(\t)\mu(d\t).
\end{equation}
Let
\be
    a_i^{v, n}\triangleq\int_{\T_i}V_i(\t)d\t+\int_{\T_0}f_i^n(\t)V_i(\t)\mu(d\t).
\ee For $f^v$ we define $a_i^v$ analogously.
From~\eqref{eq:conv_TBR} we have that $a_i^{v, n}\to a_i^v.$ Since
$\|Z_i^{v, f^n}\|_2\le 1$
for all $n,$ there exists $Z_i^v\in\Lp^2(\Omega, \Prob)$ such that
$\|Z_i^v\|_2\le 1$ and
\be
    Z_i^{v, f^n}\xrightarrow{w.}Z_i^v.
\ee
Therefore $Y_{i}^{v, n}\triangleq a_i^{v, n}\cdot Z_i^{v, f^n}$
converges weakly to
 $Y_n^v\triangleq a_i^v\cdot Z_i^v.$
It follows from the Fatou property of $\varrho_i$
that
\be
    \varrho_i(W_i-a_i^v\cdot Z_i^v)\le\liminf_{n\to\infty}\varrho_i(W_i-a_i^{v, n}\cdot Z_i^{v, f^n}).
\ee
From the lower semicontinuity of the norm in terms of weak convergence we have
\be
    \V\left[Z_i^v{V_i(\t)}\right]\le -v_i'(\t).
\ee
Proceeding as in Section~\ref{section:Min_risk_fixed_incomes}, we have that
\be
    \tilde{R}_i(v_1, v_2, f^v)\le \varrho_i(W_i-a_i^v\cdot Z_i^v).
\ee Therefore
\be
    \sum_{i=1}^2\big(\tilde{R}_i(v_1, v_2, f^v)-I_i(v_1, v_2, f^v)\big)=
    \inf_{f\in F}\Big\{\sum_{i=1}^2\big(\tilde{R}_i(v_1, v_2, f)-I_i(v_1, v_2, f)\big)\Big\}.
\ee
We denote by $({X}_1^{v}, {X}_2^{v})$ any optimal list of claims associated to the pricing schedules $(v_1, v_2)$ and the TBR $f^v.$ For notational convenience we define
\be
    \A(v_1, v_2)\triangleq\sum_{i=1}^2\big(\tilde{R}_i(v_1, v_2, f^v)-I_i(v_1, v_2, f^v)\big).
\ee
\subsubsection{Minimizing with respect to the firms' incomes}\label{subsubsec:Min_wrt_inc}
To finalize the proof of existence of a SEA, we let $\{(v_1^n,
v_2^n)\}$ be a minimizing sequence of
\be
    \sum_{i=1}^2\big(\tilde{R}_i(v_1, v_2, f^v)-I_i(v_1, v_2, f^v)\big).
\ee
We get from Proposition~\ref{prop:uniform_bound_v} that there
exist $(\bar{v}_1, \bar{v}_2)\in\C_1\times\C_2$ such that
$v_i^n\to\bar{v}_i$ uniformly. This implies that $V_i^n\to
\bar{V}_i$ almost surely (see for example Proposition A.4
in~\cite{kn:em}). Moreover, for any $\t\in\T$ where convergence
holds, it is uniform. We have from Fatou's lemma and the fact that
$0\le -I_i(v_1, v_2, f)$ for any $(v_1, v_2)\in\C_1\times\C_2$ and
any $f\in F$ that
\be
    \sum_{i=1}^2-I_i(\bar{v}_1, \bar{v}_2, f^{\bar{v}})\le
    \liminf_{n\to\infty} \sum_{i=1}^2-I_i(v_1^n, v_2^n, f^{v^n}).
\ee
In order to deal with the risky part of the firms' problems we require the following
\begin{lemma}\label{lemma:strong_weak} Let $\{\phi_n\},\,\{\psi_n\}\subset\Lp^2(\O, \Prob)$ and $\phi,\,\psi\in\Lp^2(\O, \Prob)$ be such that
\be
    \phi_n\xrightarrow{\|\cdot\|_2}\phi\quad{\mbox{and}}\quad\psi_n\xrightarrow{w.}\psi,
    \quad{\mbox{then}}\quad\langle \phi_n, \psi_n\rangle\to\langle \phi, \psi\rangle,
\ee
where $\langle\cdot, \cdot\rangle$ is the canonical inner product in $\Lp^2(\O, \Prob).$
\end{lemma}
\begin{Proof}
Adding and subtracting $\langle \phi, \psi_n\rangle$ from $|\langle \phi_n, \psi_n\rangle-\langle \phi, \psi\rangle|$ we obtain
\be
    \big|\langle \phi_n, \psi_n\rangle - \langle \phi,
    \psi\rangle\big|\le\big|\langle \phi_n-\phi, \psi_n\rangle\big|+ \big|\langle \phi,
    \psi_n-\psi\rangle\big|.
\ee
Since $\{\psi_n\}$ is a weakly convergent sequence, it is bounded.
Let $\bar{K}$ be such bound, then using the Cauchy--Schwarz
inequality we have
\be
    \big|\langle \phi_n, \psi_n\rangle - \langle \phi,
    \psi\rangle\big|\le \bar{K}\,\| \phi_n-\phi\|_{2}+ \big|\langle \phi,
    \psi_n-\psi\rangle\big|.
\ee
As $n\to\infty,$ the first summand on the righthand side of the
inequality converges to zero due to the strong  convergence of the
$\phi_n$'s to $\phi;$ the second
summand converges to zero due to the weak convergence of the
$\psi_n$'s to $\psi,$ which concludes the proof.

\end{Proof}

\noindent We show in Proposition~\ref{prop:indic_conv} that if it
were the case that the market segments exhibited no jumps in their limiting behavior then then there would exist SEAs. By absence of jumps we mean that
\be
    \i_{\Theta_0^n}\xrightarrow{a.s.}\i_{\bar{\Theta}_0}\quad{\mbox{and}}\quad\i_{\Theta_i^n}\xrightarrow{a.s.}\i_{\bar{\Theta}_i},
\ee
where  $\bar{\Theta}_0 = \{\bar{v}_1 = \bar{v}_2\}$ and $\bar{\Theta}_i= \{\bar{v}_i > \bar{v}_{-i}\}.$ This ``nice'' convergence of the $\T_i^n$'s is by no means the general scenario. As an example let $v_1^n\equiv v_1$ and $v_2^n=v_1+1/n.$ Here $\T_1^n\equiv \T,$ $\T_2\equiv\emptyset,$ but in the limit only $\bar{\Theta}_0\neq\emptyset.$ Nonetheless, the existence result of SEAs under the assumption of convergence of the indicator functions is relevant for the general case, and we present it below.

\begin{proposition}\label{prop:indic_conv} Let $\{(v_1^n, v_2^n)\},$ $(\bar{v}_1, \bar{v}_2)\subset\C_1\times\C_2$ such that $v_i^n\to\bar{v}_i$ uniformly.
Assume that  $\i_{\Theta^n_j} \to \i_{\bar{\Theta}_j},$ and
let $\bar{X}_1, \bar{X}_2$ and $\bar{f}$ be the (weak) limits of $X^{v^n}_1, X^{v^n}_2$
and $f^{v^n}.$ Then
\begin{eqnarray*}
    \liminf_{n\to\infty} \sum_{i=1}^2\tilde{R}_i(v_1^n, v_2^n, f^{v^n}) & \geq & \sum_{i=1}^2
    \varrho_i\left(W_i - \int_{\bar{\T}_i} \bar{X}_i \mu(d \t) -
    \int_{\bar{\T}_0} \bar{f}_i\, \bar{X}_i \mu(d \t) \right) \\
    & \geq & \sum_{i=1}^2 \varrho_i\left(W_i - \int_{\bar{\T}_i} X^{\bar{v}}_i \mu(d \t) -
    \int_{\bar{\T}_0} f^{\bar{v}}_i X^{\bar{v}}_i \mu(d \t) \right),
\end{eqnarray*}
where $(X^{\bar{v}}_1, X^{\bar{v}}_2, f^{\bar{v}})$ solves the social planer's problem for
$(\bar{v}_1, \bar{v}_2)$.
\end{proposition}
\begin{Proof} The second inequality follows from the definition of $(\bar{X}_1, \bar{X}_2, \bar{f}).$ To
show that the first one holds, we first use Proposition~\ref{prop:representation_optimal_contracts} and write
\be
    \int_{\T_i^n} X_i^{v^n, f^{v^n}} \mu(d\t) - \int_{\T_0^n} f_i^n\, X_i^{v^n, f^{v^n}} \mu(d \theta)=
    Z_i^{v^n, f^{v^n}}\int_{\T} \Big(\i_{\T_i^n}(\t)- f_i^{v^n}\, \i_{\T_0^n}\,V_i^n(\t)\Big)\mu(d \t)
    \triangleq a_i^{v^n}\cdot Z_i^{v^n, f^{v^n}}.
\ee
The assumption on the
convergence of the indicator  functions $\i_{\Theta^n_j}$ implies that
\be
    V_i^n\, \i_{\T_0^n} \xrightarrow{a.s.} \bar{V}_i\, \i_{\bar{\T}_0}.
\ee
Since $|V_i^n\, \i_{\T_0^n}|, |\bar{V}_i\, \i_{\bar{\T}_0}|\le M$ and $\mu(\O)<\infty,$ by Lebesgue Dominated Convergence we get
\be
    V_i^n\,\i_{\T_0^n}\xrightarrow{\|\cdot\|_2}\bar{V}_i\, \i_{\bar{\T}_0}.
\ee
Hence, by Lemma~\ref{lemma:strong_weak} we have that $a_i^{v^n}\to\bar{a}_i,$
where
\be
    \bar{a}_i\triangleq\int_{\Theta} \Big(\i_{\bar{\T}_1}(\t)- \bar{f}\, \i_{\bar{\T}_0}\,\bar{V}_i(\t)\Big)\mu(d \theta).
\ee
Therefore
\be
    a_i^{v^n}\cdot Z_i^{v^n}\xrightarrow{w.} \bar{a}_i\cdot \bar{Z}_i,
\ee
and the Fatou property of the risk measure yields
\be
    \sum_{i=1}^2{R}_i(\bar{v}_1, \bar{v}_2, {X}_i, {f})\le\liminf_{n\to\infty} \sum_{i=1}^2\tilde{R}_i(v_1^n, v_2^n, f^{v^n}).
\ee

\end{Proof}

\noindent We now deal with the possibility of non--convergence of
the indicator functions. We first observe that the $\T^n_0$'s are
closed subsets of $\T,$ which is compact. The set \be
    \bar{2}^{\T}:=\{A\subset\T\,\mid\,A\neq\emptyset\,\,{\mbox{and $A$ is closed}}\}
\ee
endowed with the Hausdorff metric $h:\bar{2}^{\T}\to\re_+$ is a compact metric space (see for example~\cite{kn:ab}, Chapter 3). If $\T^n_{0}\neq\emptyset$ infinitely often, then there exists $\hat{\T}_{0}\in\bar{2}^{\T}$ such that, up to a subsequence if necessary
\be
    \T^n_{0}\xrightarrow{h}\hat{\T}_{0}.
\ee
Moreover, $\hat{\T}_{0}$ is contained in $\bar{\T}_{0}.$ If, on the contrary, $\T^n_{0}=\emptyset$ for all but a finite number of $n$'s, then we define the Hausdorff limit of $\{\T^n_{0}\}$ as the empty set, which is again contained in $\bar{\T}_{0}.$
In both instances we observe that there can only be more ties in the limit, and
hence more ways of breaking them. This suggests that the aggregate
risk in the limit is indeed no greater than the aggregate risk in
the pre--limit.

As for the sets $\T_i^n,$ if $\theta$ is not eventually in
either of them, then either $\theta\in \hat{\T}_{0}$ or it is a type
whose preferences alternate. However, the limiting behavior of the
latter is that of indifference, due to the convergence of the
functions $v^n_i$. In other words, an agent type $\t$
eventually always contracts with the same firm $i$, alternates
between firms or is always indifferent. Thus,
\be
    \t \in \T^n_0 \,\, \mbox{for all sufficiently large $n$ or} \,\, \t \in \bar{\T}_0.
\ee
We now decompose the type space into subsets for which
the associated indicator functions converge. To this end, we define
\be
    \tilde{\T}^n_0 \triangleq \bar{\T}_{0}\setminus\T_0^n, \quad{\mbox{and}}\quad \tilde{\T}^n_i \triangleq \T_i^n\setminus \tilde{\T}^n_0.
\ee
The sets $\tilde{\T}^n_i$ contain the ``surviving customers'',
in the sense that there will be no jumps towards indifference from
types in $\tilde{\T}^n_i$ at the limit; $\tilde{\T}^n_0$ is the set
of ``alternating customers''.
By construction \be
    \T= \tilde{\T}^n_0 \cup \T^n_0\cup \tilde{\T}^n_1 \cup \tilde{\T}^n_2
    = \bar{\T}_{0} \cup \tilde{\T}^n_1 \cup \tilde{\T}^n_2.
\ee  The following lemma shows that the sets $\tilde{\T}^n_i$ of
``surviving customers'' converge to the sets of agent types that
contract with firm $i$ when $(\bar{v}_1, \bar{v}_2)$ is offered.

\begin{lemma}\label{lm:Conv-Limit-No-ties}
For $i=1,2$ we have $\i_{\tilde{\T}^n_i} \to {\i}_{\bar{\T}_i}.$
\end{lemma}
\begin{Proof}
Let $\t \in \T$ be given. There are two possible cases.
Either there exists $N\in\n$ such that if $n>N,$ then $\t  \in
\tilde{\T}^n_i$ or there exists a subsequence $\{v_1^{n_k},
v_2^{n_k}\}$ such that $\t\notin \tilde{\T}^{n_k}_i$ for all $n_k$.
In the former case the (uniform) convergence of the $v_i^n$'s
implies $\bar{v}_i(\t) > \bar{v}_{-i}(\t)$. In the latter
case, either $\t$ eventually belongs to $\tilde{\T}^n_{-i},$ in
which case convergence of the indicator functions follows or $\t\in
\tilde{\T}^n_1 \cap \tilde{\T}^n_2$ infinitely often. In this case
$\t\in\tilde{\T}^n_0$ and we again have convergence of the
indicators.

\end{Proof}

\noindent Let us now assume that the social planer's problem were
such that at every stage $n$ all agent types that belong to
the set of ``alternating customers'' are deemed indifferent. This
results in more tie--breaking possibilities. Specifically we
consider the social planer's problem
\begin{small}
\be
    \A^*(v^n_1, v^n_2) = \inf_{f^n\in F} \inf_X\bigg\{\sum_{i=1}^2
    \varrho_i\Big(W_i - \int_{\tilde{\T}^n_i} X^n_i \mu(d \theta) -
    \int_{\tilde{\T}^n_0\cup\T^n_0} f^n_i X^n_i \mu(d \theta) \Big)\Big| \V[X^n_i(\t)]=-(v_i^n)'(\t)\bigg\}.
\ee
\end{small}
A subset of the possible choices of the $f^n_i$'s is $F_0^n,$ which is defined as the set of $\tilde{f}\in F$ such that
\be
\begin{array}{cc}
  \tilde{f}(\t)= & \left\{
        \begin{array}{ll}
          0, & v_i^n(\t)<v_{-i}^n(\t) \\
          1, & v_i^n(\t)>v_{-i}^n(\t) \\
          f(\t), & \hbox{otherwise},
        \end{array}
      \right.
\end{array}
\ee
for some TBR $f\in F.$ We then have that
\be
    \A^*(v^n_1, v^n_2)|_{F_0}=\A(v^n_1, v^n_2).
\ee
Thus,
\be
    \A(v^n_1, v^n_2) \geq \A^*(v^n_1, v^n_2).
\ee
Consider a minimizing sequence $\{(v^n_1, v^n_2)\}$ with
(uniform) limit $(\bar{v}_1, \bar{v}_2).$ Lemma~\ref{lm:Conv-Limit-No-ties} guarantees that $\i_{\tilde{\T}^n_i}\to {\i}_{\bar{\T}_i};$ moreover $\tilde{\T}^n_0\cup\T^n_0\equiv \bar{\T}_{0}$ so patently $\i_{\tilde{\T}^n_0\cup\T^n_0}\to\i_{\bar{\T}_{0}}.$ These two facts  allow us to apply Proposition~\ref{prop:indic_conv} to $\A^*(v^n_1, v^n_2),$ which yields
\begin{eqnarray*}
    \liminf_{n\to\infty} \A(v^n_1, v^n_2) &\geq& \liminf_{n\to\infty} \A^*(v^n_1, v^n_2)
    \\
    & \geq & \sum_{i=1}^2 \varrho_i\bigg(W_i - \int_{\bar{\T}_i} X^*_i \mu(d \theta) -
    \int_{\bar{\Theta}_0} f^*_i X^*_i \mu(d \theta) \bigg),
\end{eqnarray*}
where $(X^*_1, X^*_2, f^*)$ solves the social planer's problem for
$(\bar{v}_1, \bar{v}_2)$. This shows that $(X^*_1, X^*_2, f^*)$ is indeed
optimal because $\{(v^n_1, v^n_2)\}$ was required to be a minimizing sequence. We have proved the following

\begin{thm}\label{theorem:Main}
If firm $i$ assesses risk using a law invariant risk measure
\be
    \varrho_i:\X\to\re\cup\{+\infty\},
\ee
which has the Fatou property, and if it offers catalogues of the
form
\be
    C_i=\big\{(X_i(\t), p_i(\t))\big\}_{\t\in\T},
\ee
then there exists a socially efficient market situation.
\end{thm}

\subsubsection{Individual rationality revisited}\label{subsubsec:Ind_Rat_revisited}

We can also deal with a slightly different setting, in which the
regulator's presence would be necessary for the enforcement of a
cash transfer. Namely, assume that $(C_1^*, C_2^*, f^*)$ is a
solution to the problem $\Psi$ without the individual
rationality constraints  $\R_i(C_1, C_2, f)\le\varrho_i(W_i).$
Since both firms could simply offer $(0, 0),$ a minimization of
$\A(\cdot,\cdot,\cdot)$ would be IR on the aggregate level. Then
there would exist $r\ge 0$ (the {\sl rent of risk exchange}) such
that
\[
A(C_1^*, C_2^*, f^*)=\varrho_1(W_1)+\varrho_2(W_2)-r
\]
We define a {\sl transfer SEA} to be a quadruple $(C_1^*, C_2^*,
f^*, T^*)$ such that
\begin{enumerate}

\item $\Psi=A(C_1^*, C_2^*, f^*),$

\item $\R_1(C_1^*, C_2^*, f^*)-T^*\le\varrho_1(W_1),$

\item $\R_2(C_1^*, C_2^*, f^*)+T^*\le\varrho_2(W_2),$
\end{enumerate}
and $T^*\in[r_1, r_2]$ for some $r_1, r_2\in\re$ such that
$r_2-r_1=r.$ The arguments contained in
Sections~\ref{section:Min_risk_fixed_incomes},
\ref{subsubsection:ETR_Risk} and~\ref{subsubsec:Min_wrt_inc} can
be immediately applied to prove the following
\begin{corollary}
If firm $i$ assesses risk using a law invariant risk measure
\be
    \varrho_i:\X\to\re\cup\{+\infty\},
\ee
which has the Fatou property, and if it offers catalogues of the
form
\be
    C_i=\big\{(X_i(\t), p_i(\t))\big\}_{\t\in\T},
\ee
then there exists a transfer socially
efficient market situation.
\end{corollary}

\section{Examples: Risk Minimization}\label{sec:examples}

In this section we focus our attention on two well--known risk
measures: entropic and average value at risk. When it comes to the
latter, we mostly present in Section~\ref{subsec:simul_AVAR} the
results of applying the numerical algorithm that is described in
Appendix~\ref{app:Algorithm_risk} to some specific
examples\footnote{All of our codes are available upon request}. As
for the entropic risk measure, before proceeding with the numerical
simulations, we present in Section~\ref{subsubsec:Entropic} a
structural result that exploits the risk measure's particular
structure. Moreover, we show that in this case there is a SEA where
both firms service all of the agents, and for which the optimal TBR
is a constant proportion over the whole market. We refer to such
efficient TBRs as  ``fix--mix'' rules.

\subsection{Entropic--risk--minimizing firms}\label{subsubsec:Entropic}

In what follows we concentrate on a situation where the firms use the entropic risk measure as a means to assess their risk exposure, i.e.
\be
\varrho_i(X)=\frac{1}{\gamma_i} \ln \E_\Prob \left[\exp(-\gamma_i X)\right].
\ee
The coefficient $\gamma_i$ represents firm $i$'s risk aversion. This particular choice of risk measures, which is closely related to exponential utility, allows us to further the analysis into the structure of SEAs. Moreover, given that it is strictly convex and that it has a closed--form representation it lends itself very well to numerical exercises. For $v_i$ and $f$ given, Proposition~\ref{prop:representation_optimal_contracts} allows us to write program $\Psi_1$ (the minimization with respect to the firms' claims for fixed incomes and TBR in Section~\ref{section:Min_risk_fixed_incomes}) as
\be
\inf_{(Z_1, Z_2)\in\Gamma_i\times\Gamma_2}\sum_{i=1}^2\varrho\bigg(W-Z_i\Big(\int_{\T_i}\sqrt{-v_i'(\t)}\mu(d\t)+\int_{\T_0}\sqrt{-v_i'(\t)}f_i(\t)\mu(d\t)\Big)\bigg),
\ee
where
\be
    \Gamma_i:=\{Z\in\X_i\,\mid\, \E[Z]=0,\,\|Z\|^2_2=1\}.
\ee
The structure above allows us to write (See Section 3.4.2 in~\cite{kn:hm}) the minimization problem of firm $i$ as that of finding a stationary point to the Lagrangian
\be
L(Z,\tau,\kappa)=\ln\left(\int_{\O}e^{-\gamma_i(W+Z
a(v_i, f_i))}d \Prob\right)+\tau\int_{\O}Z d \Prob+\kappa\int_{\O}Z^2 d\Prob,
\ee
where $\tau$ and $\kappa$ are the Lagrange multipliers associated to the moment constraints, and
\be
a(v_i, f_i)=\int_{\T_i}\sqrt{-v_i'(\t)}\mu(d\t)+\int_{\T_0}\sqrt{-v_i'(\t)}f_i(\t)\mu(d\t).
\ee
This in turn results in the following implicit representation for the optimal claims given $v_i$ and $f,$ which has a unique solution for each realization $z_i^{v, f}$ of $Z_i^{v, f}$ and $w_i$ of $W_i:$
\begin{equation}\label{eq:implicit_entropic}
Z_i^{v, f}=-\frac{e^{-\gamma_i(W_i+Z_i^{v, f} a(v_i, f_i))}-\E\left[e^{-\gamma_i(W_i+Z_i^{v, f} a(v_i, f_i))}\right]}{\sqrt{\V\left(e^{-\gamma_i(W_i+Z_i^{v, f} a(v_i, f_i))}\right)}}.
\end{equation}
\begin{remark}\label{rem:aggregator}
The indirect utility functions $v_i$ and the TBR $f_i$ only affect the optimal claims $Z_i^{v, f}$ via the ``aggregator''  $a(v_i, f_i).$
\end{remark}
The previous remark has interesting repercussions for the socially efficient TBRs. Indeed, assume the firms have initial endowments $W_i$ and they are characterized by risk aversion coefficients $\gamma_i.$ Theorem~\ref{theorem:Main} guarantees the existence of a SEA $(v_1^*, v_2^*, f^*, Z_1^*, Z_2^*),$ which yields the following aggregate risk in the economy:
\begin{equation}\label{eq:aggregate_risk_entropic}
\varrho_1\big(W_1-a(v_1^*, f_1^*)\,Z_1^*\big)+\varrho_2\big(W_2-a(v_2^*, f_2^*)\,Z_2^*\big)+I[v_1^*, v_2^*],
\end{equation}
where $I[v_1^*, v_2^*]$ represents the firms' aggregate income (which is independent of $f^*$). We know from Remark~\ref{rem:aggregator} that any modification on $f^*$ that leaves  $a(v_i^*, f_i^*)$ unchanged bears no weight on the value of expression~\eqref{eq:aggregate_risk_entropic}. We can redefine
\be
\begin{array}{cc}
  f^*(\t)= & \left\{
        \begin{array}{ll}
          0, & v_1(\t)<v_2(\t) \\
          1, & v_1(\t)>v_2(\t) \\
          f^*(\t), & \hbox{otherwise},
        \end{array}
      \right.
\end{array}
\ee
and define
\be
v^*(\t)\triangleq\max\big\{v_1^*(\t), v_2^*(\t)\big\},
\ee
which then allows us to write
\be
a(v_i^*, f_i^*)=\int_{\T}\sqrt{-(v^*)'(\t)}f_i^*(\t)\mu(d\t).
\ee
Once we have defined the TBR over the whole $\T,$ we may go one step further and write
\be
K\triangleq\dfrac{\int_{\t}\sqrt{-(v^*)'(\t)}f^*(\t)\mu(d\t)}{\int_{\t}\sqrt{-(v^*)'(\t)}\mu(d\t)},
\ee
then
\be
a(v_1^*, f_1^*)=K\int_{\T}\sqrt{-(v^*)'(\t)}\mu(d\t)\quad{\text{and}}\quad a(v_2^*, f_2^*)=(1-K)\int_{\T}\sqrt{-(v^*)'(\t)}\mu(d\t).
\ee
An interesting economical conclusion from the computations above is that there exists a SEA that consists of both firms servicing the whole market, and splitting the customers in a $K$ to $1-K$ proportion. This follows from the fact that only $v^*$ (the upper envelope of the original $v_i^*$) appears in each firm's program. In markets such as regulated health--insurance, where firms are legally prevented from abstaining from contracting with any agent, the regulator may oversee that a socially optimal proportion of the market is serviced by each firm.

\subsection{Simulations (Entropic--risk minimizing firms)}\label{subsec:sim_entropic}

In this section we provide the numerical analysis of a particular
example of the entropic--risk minimizing firms analyzed above.  To
this end we set
\begin{itemize}
\item Dimension of the space defining $v$: $6,$ dimension of $\Omega$: $14,$
\item $W_1=0.5*(-1, -3, -9, -3, -1, -0.2, -0.1, -0.1, -0.2, 1, -3, -9, -3, -1)^T,$
\item $W_2=0.5*(-0.03, -0.1, -0.18, -0.2, -1, -3, -9, -10, -3, -1, -0.2, -0.18, -0.1, -0.03)^T,$
\item risk aversion coefficient $\gamma=2.$
\end{itemize}

\subsubsection{Risk assessments}

\begin{figure}[ht!]
\begin{center}
\subfigure[\label{fig:Entropic_comp} Numerical v.s. theoretical
values for $Z$]
{\includegraphics[width=0.48\textwidth]{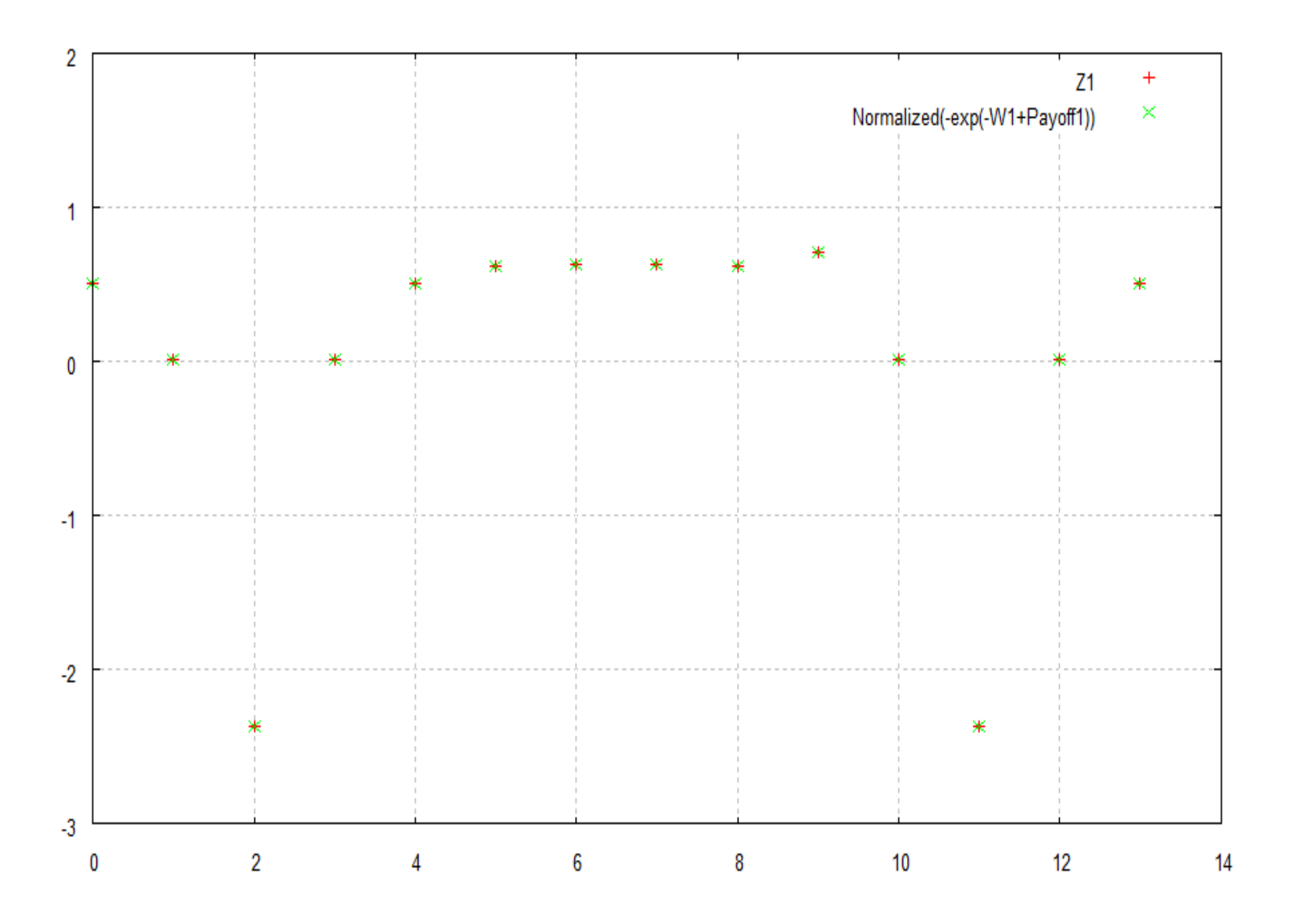}}\hspace{.2in}
\subfigure[\label{fig:Entropic_v}The minimizing $v$ ]
{\includegraphics[width=0.48\textwidth]{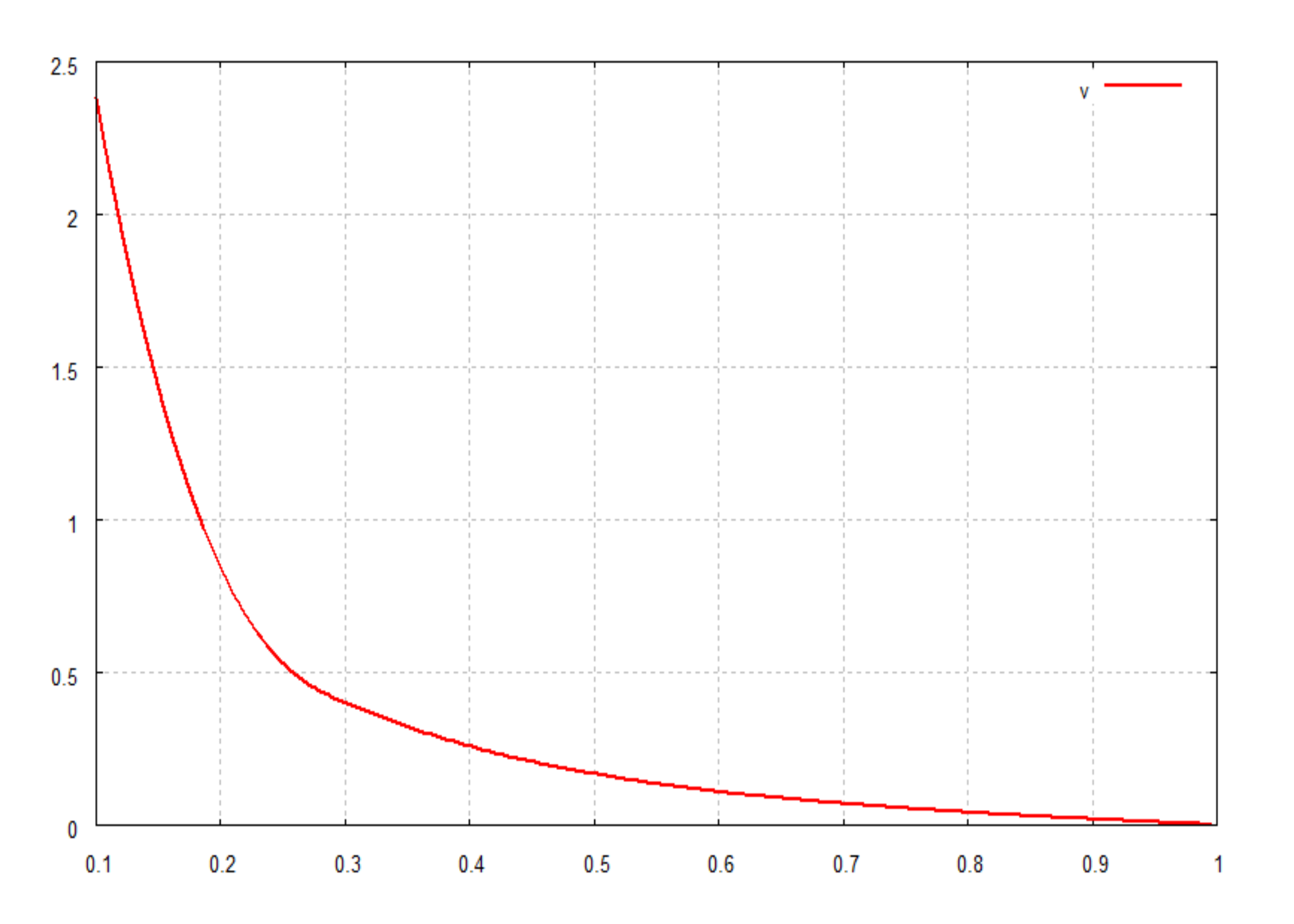}}
\caption{\label{figure:1} Comparison results for the $f\equiv 1$
case.}
\end{center}
\end{figure}

Let us benchmark the aggregate risk in the competitive economy
against the risk in a monopolistic setting. To this end, we first
fix $f\equiv 1$. This corresponds to a model in which firm 1 acts as
a monopolist and firm 2 has no access to the market. The a-priori
aggregate risk in the economy is $7.36$. The risk assessment of firm
1 is reduced from $3.53$ to $2.16$ after the it has traded with the
agents while the risk of firm 2 stays the same. Below we plot the
numerical result for $Z,$ as well as the theoretical one from
Equation~\eqref{eq:implicit_entropic} in
Figure~\ref{fig:Entropic_comp} and the corresponding minimizing $v$
(the agents' indirect utility) in Figure~\ref{fig:Entropic_v}.
Likewise, if we fix $f\equiv 0,$ which again yields a
principal--agent setting (competition--wise) for firm 2, we obtain
that this firm's initial risk is $3.84,$ which is reduced to $2.30$
after trading. Finally, once we let $f$ vary, the aggregate risk
decreases from $7.36$ to  $5.39,$ and the corresponding final risk
assessments for the firms are $2.17$ and $2.67$ respectively. While
each individual firm is worse off in the presence of competition, it
should be noted that the decrement in the aggregate risk in the
economy is lower in either case where only one firm has access to
the market. Risk decreases from $7.36$ to $5.99$ when only firm 1 is
active, and from $7.36$ to $5.72$ if it is firm 2 who trades with
the agents.

\subsubsection{Risk profiles}

\begin{figure}[ht!]
\begin{center}
\subfigure[\label{fig:Entropic_before_after_1} Comparison of $W_1$ and $W_1-a(v_1^*, f_1^*)Z_1^*$]
{\includegraphics[width=0.49\textwidth]{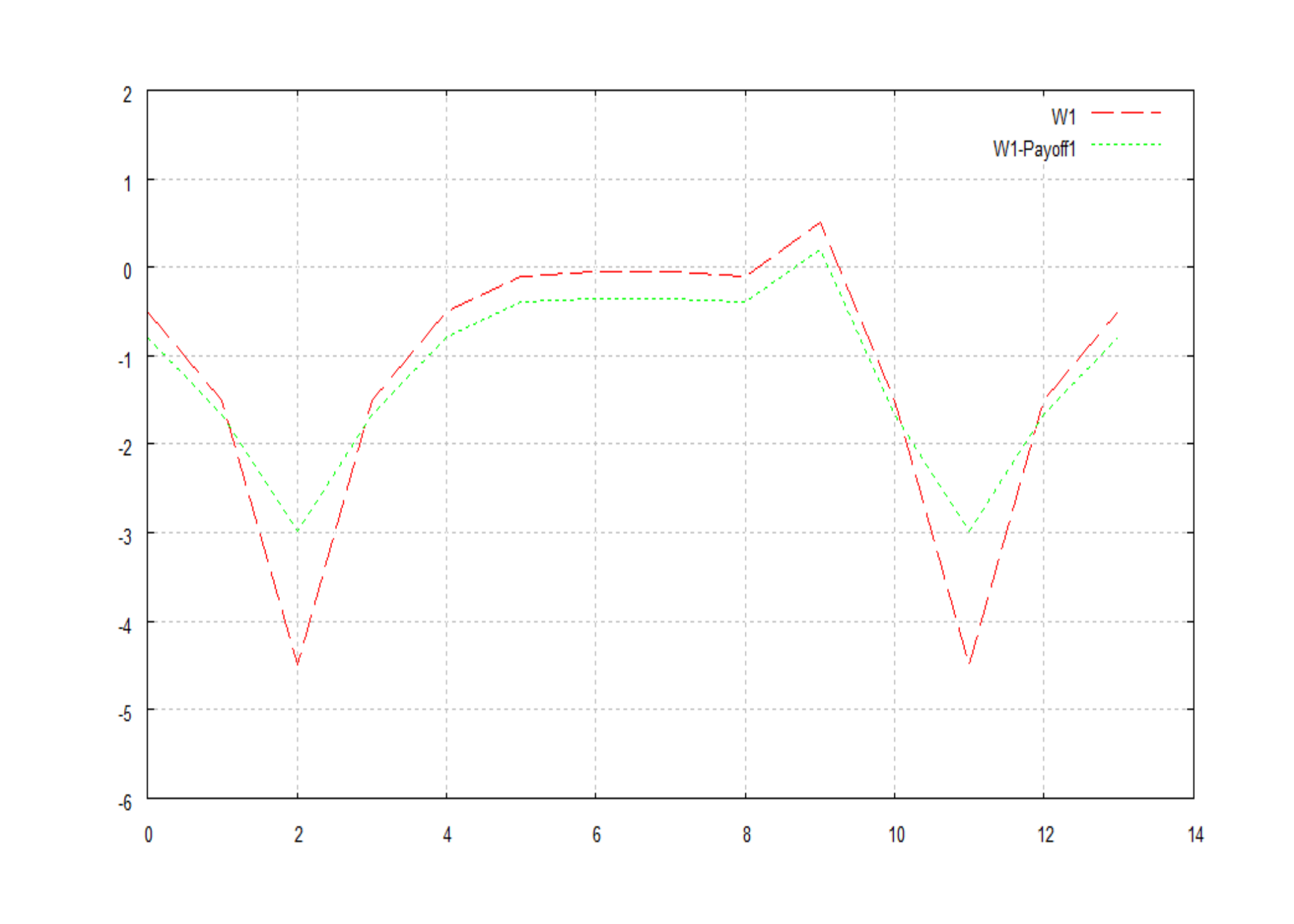}}
\subfigure[\label{fig:Entropic_before_after_2}Comparison of $W_2$ and $W_2-a(v_2^*, f_2^*)Z_2^*$]
{\includegraphics[width=0.49\textwidth]{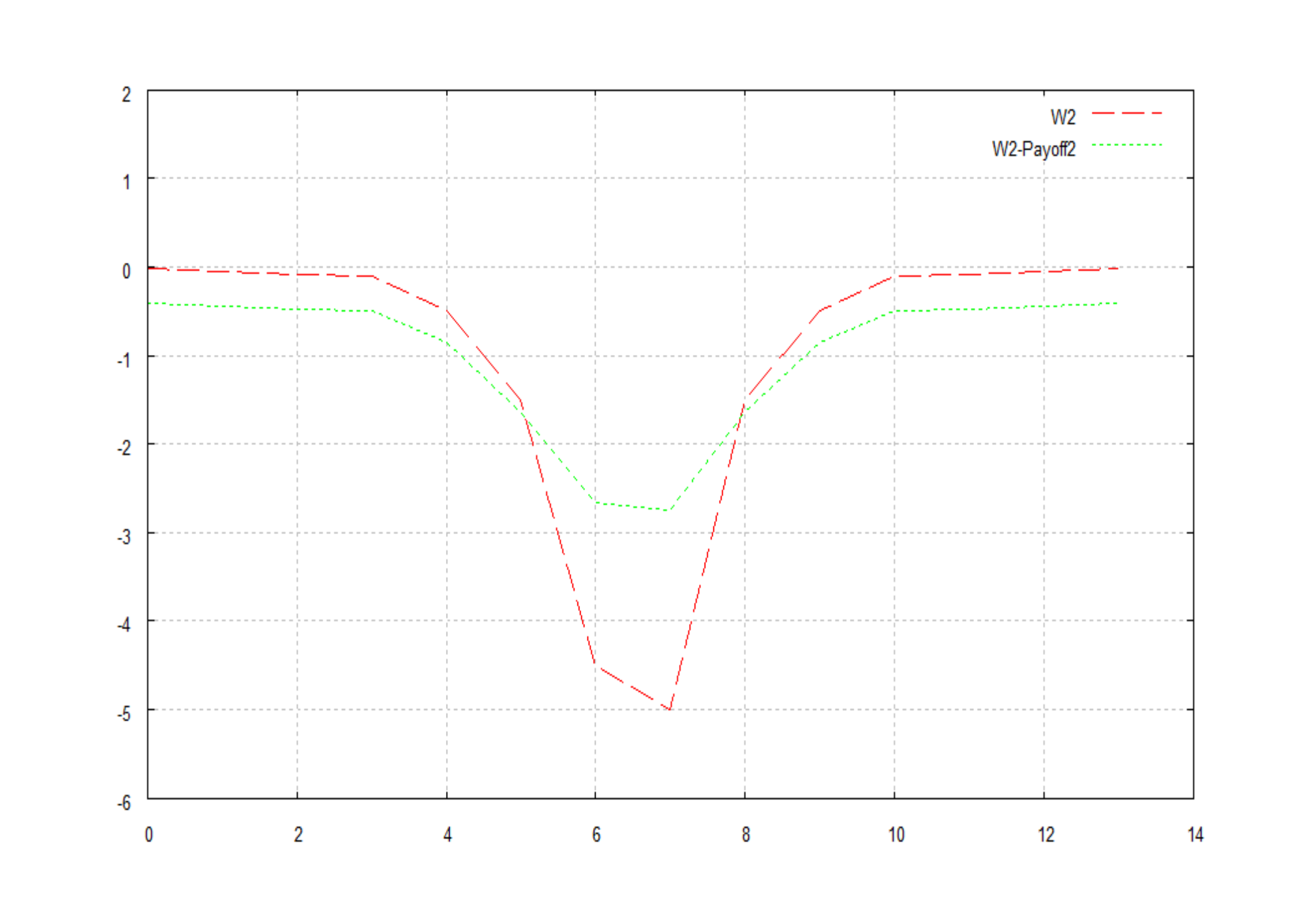}}
\caption{\label{figure:2}
Positions before and after trading.}
\end{center}
\end{figure}

\begin{figure}[ht!]
\begin{center}
{\includegraphics[width=0.6\textwidth]{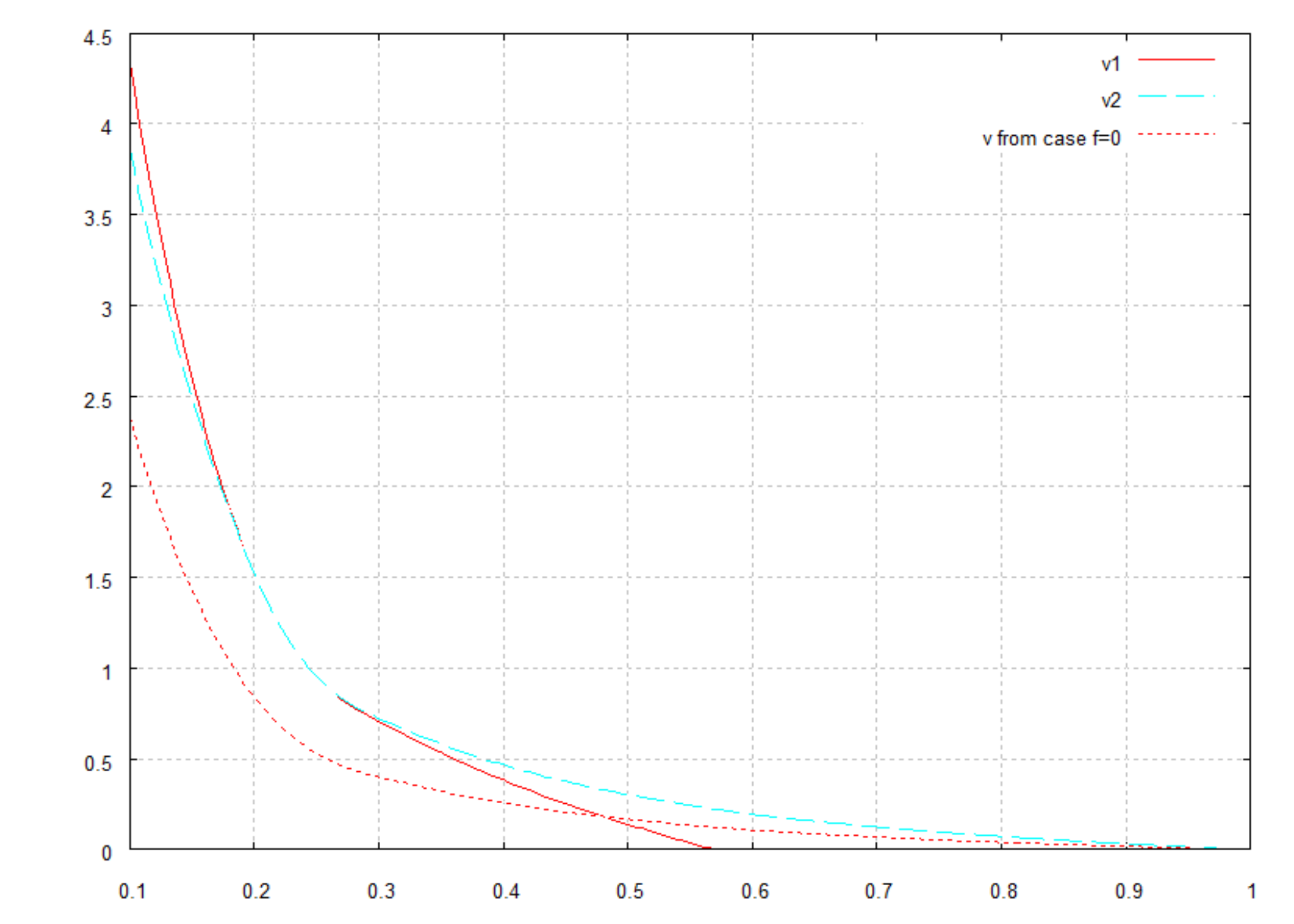}}
\caption{\label{figure:3}
Indirect utilities under monopoly and oligopoly.}
\end{center}
\end{figure}

With respect to the ex--ante and ex--post risk profiles, we observe
that trading has a smoothing effect, flattening spikes that
correspond to the bad states of the World. However the basic shape
of the risk profile remains, which is in contrast to what we find in
Figure~\ref{figure:4} of our AV@R example. This is presented in
Figure~\ref{figure:2}, where the $14$ elementary events have been
connected by lines for illustration purposes.
Figure~\ref{figure:3} shows the agents' indirect utilities
associated to each firm's offer, and compares it to the indirect
utility for the agents who face a monopolist (in this case firm 1).
The market is shared at a $0.42$ to $0.58$ ratio between the firms.
We observe that the upper envelope $u(\t)=\max\{v_1(\t), v_2(\t)\}$
dominates the monopolistic situation for all agents. In conclusion,
this particular example is in line with the intuition that
competition among sellers benefits the buyers. Moreover, the
competitive setting also provides a lower aggregate risk exposure. A
point could be made that the regulator should make sure that enough
incentives exist for all firms to engage in risk--minimizing
trading, as it is socially desirable.

\begin{remark} Arguably, the implementation of an efficient TBR in full generality could prove to be a daunting task. However,
the structure of  $Z_i^{v, f}$ presented in Equation~\eqref{eq:implicit_entropic} suggests that in general these variables depend on $v$ and $f$ only via the integral expression $a(v, f).$ If such were the case, one could achieve an optimum by choosing an efficient ``fix-mix'' TBR, where each firm caters to a constant proportion of the whole market. The latter clearly makes the implementation considerably simpler. For the special case of the entropic measure this can indeed be achieved, as shown above.
\end{remark}

\subsection{Simulations (AV@R--minimizing firms)}\label{subsec:simul_AVAR}

\begin{figure}[ht!]
\begin{center}
\subfigure[\label{fig:AVAR_comp1} Comparison of $W_1$ and $W_1-a(v_1^*, f_1^*)Z_1^*$]
{\includegraphics[width=0.47\textwidth]{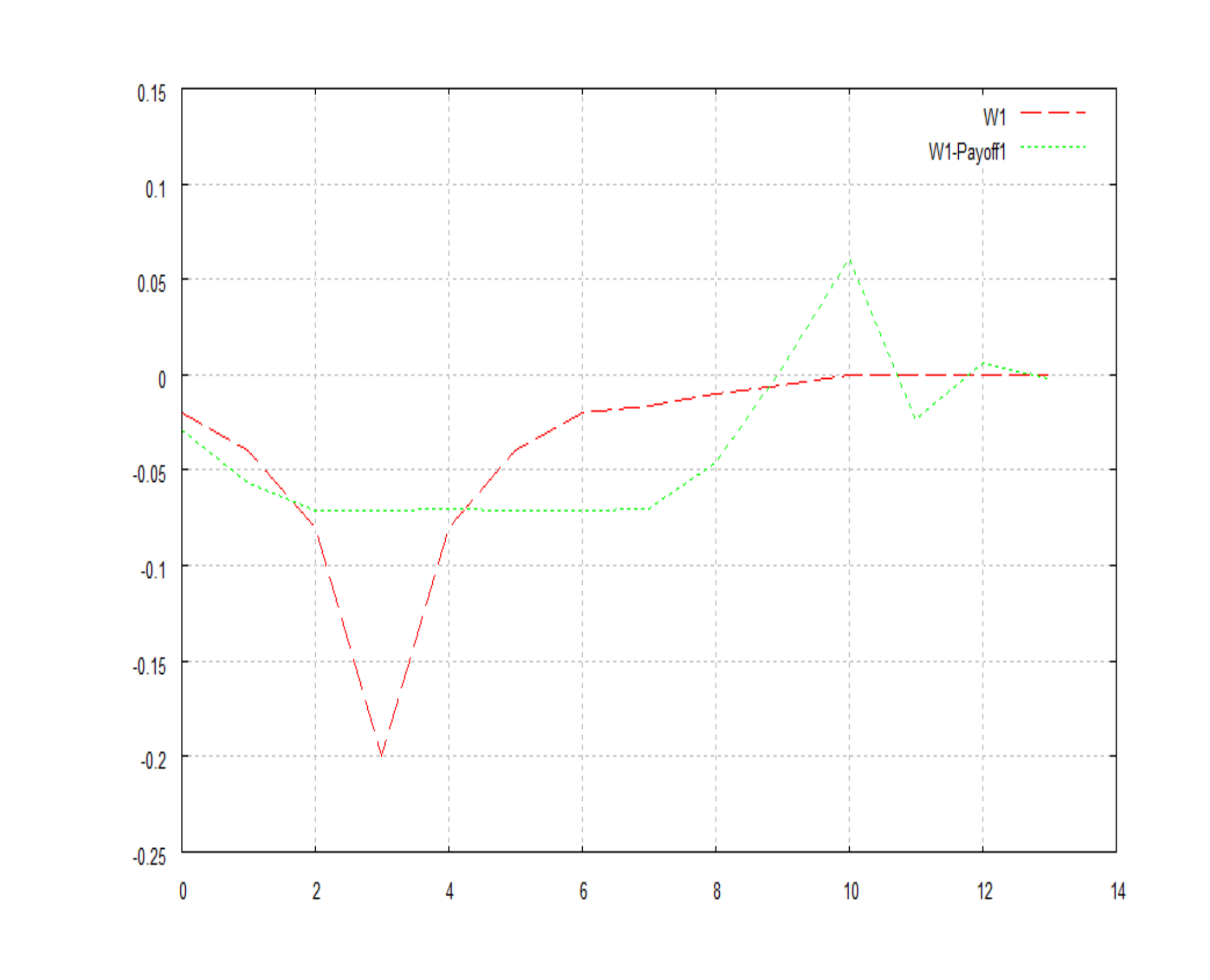}}
\hspace{.2in}
\subfigure[\label{fig:AVAR_comp2} Comparison of $W_2$ and $W_2-a(v_2^*, f_2^*)Z_2^*$]
{\includegraphics[width=0.47\textwidth]{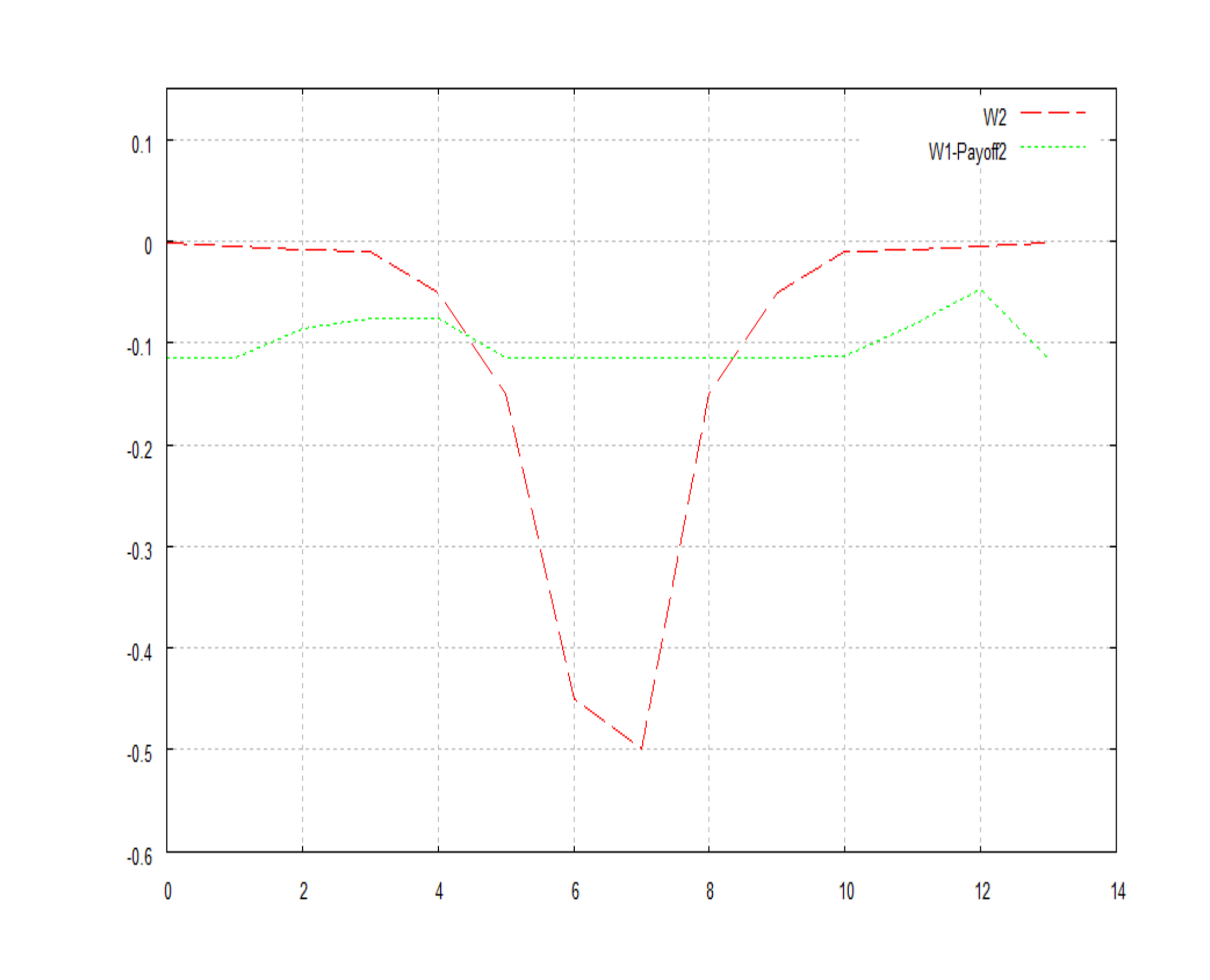}}
\caption{\label{figure:4}
Positions before and after trading.}
\end{center}
\end{figure}

\begin{figure}[ht!]
\begin{center}
\subfigure[\label{fig:AVAR_ind_uts} The indirect utility functions]
{\includegraphics[width=0.47\textwidth]{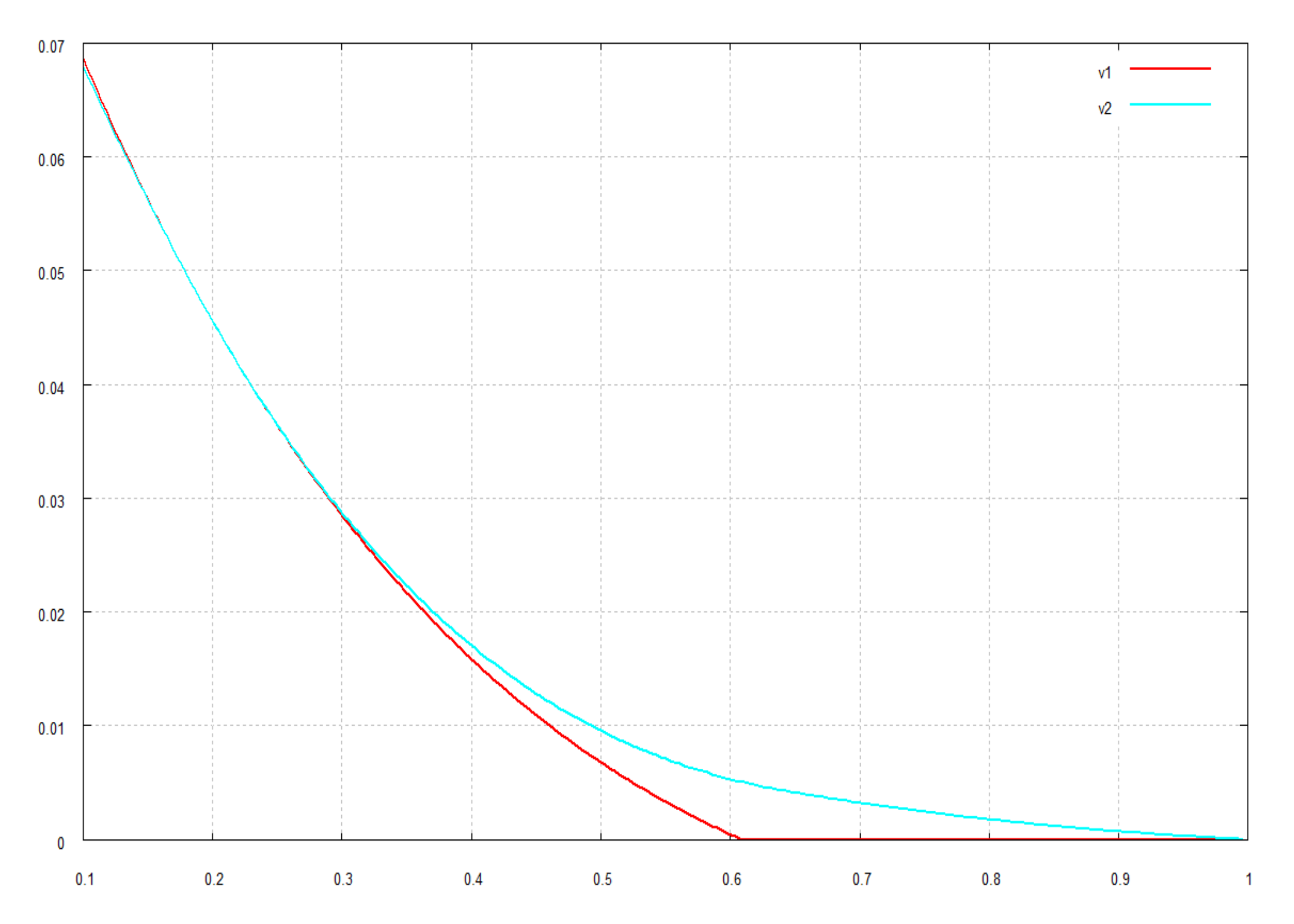}}
\hspace{.2in}
\subfigure[\label{fig:AVAR_efficint_TBR} A possible socially efficient TBR]
{\includegraphics[width=0.47\textwidth]{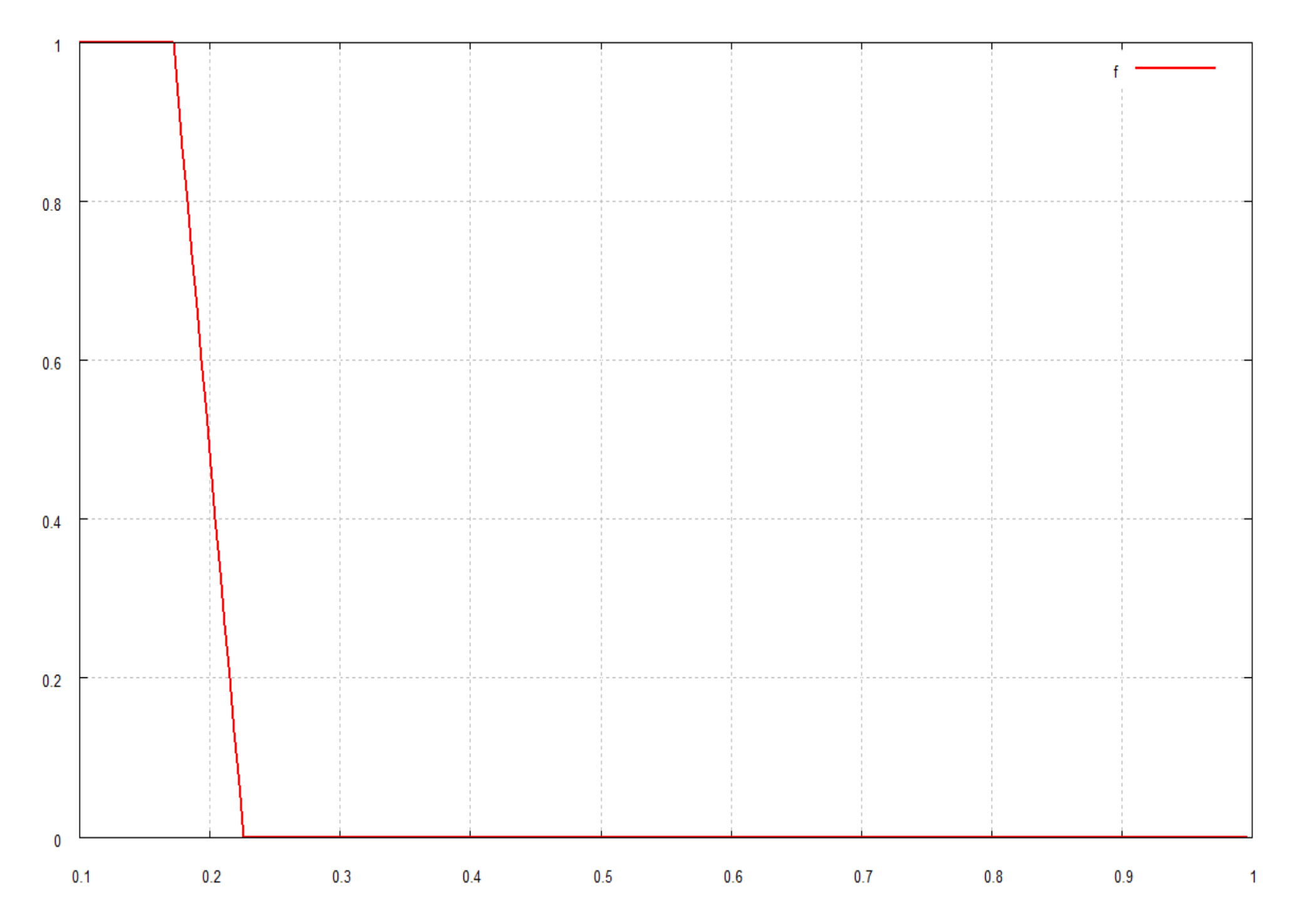}}
\caption{\label{figure:5}
Indirect utilities and a possible efficient TBR.}
\end{center}
\end{figure}

In this section we study an example where the firms are average value--at--risk minimizers. Recall that for $\lambda\in (0, 1],$ and $X\in\Lp^{\infty}(\O, \F, P),$ one defines
\be
AV@R_{\lambda}(X)\triangleq\frac{1}{\lambda}\int_0^{\lambda}V@R_{\gamma}(X)d\gamma,
\ee
where
\be
V@R_{\gamma}(X)\triangleq\inf\big\{m\,\mid\,P\{X+m<0\}\le\gamma\big\}.
\ee
We use the following initial parameters:
\begin{itemize}
\item Dimension of the space defining $v$: $6,$ dimension of $(\Omega)$: $14$,
\item $W_1=0.02*(-1, -2, -4, -10, -4, -2, -1, -0.8, -0.5, -0.3, 0, 0 , 0, 0)^T,$
\item $W_2=0.05*(-0.03, -0.1, -0.18, -0.2, -1, -3, -9, -10, -3, -1, -0.2, -0.18, -0.1, -0.03)^T,$
\item levels for the AV@R: $\lambda_1$: $0.05$ and $\lambda_2$: $0.1$
\end{itemize}
The initial aggregate initial risk assessment is $0.68,$ which decreases to $0.21$ after trading. In Figure~\ref{figure:4} we compare the firms' positions before and after trading. We observe that in contrast with the entropic--risk--measure case (see Figure~\ref{figure:2}), the ex--post shapes of the risk profiles have been significantly altered. This is due to the fact that the risk measure in hand places a heavier weight on the bad states of nature, even at the cost of the originally good ones Figure~\ref{fig:AVAR_ind_uts} shows the indirect utility functions corresponding to the catalogue that each firm offers, and in
Figure~\ref{fig:AVAR_efficint_TBR} we have plotted an efficient TBR.

\begin{remark} It should be noted that in neither of the examples presented above were the individual rationality constraints for the firms binding. This is essential to implement the ``fix--mix'' TBR without the need of a cash transfer. Although $f$ plays no role in the aggregate income $I[v_1, v_2],$ it does enter the IR constraints of each firm. If one wanted to implement a ``fix--mix'' TBR without a cash transfer scheme, they would in general run into a two--equations--one--unknown issue as soon as an IR constraint bound. \end{remark}

\section{Conclusions}

In this paper we have extended the principal--agent models of profit maximization found in~\cite{kn:cet} and risk minimization found in~\cite{kn:hm} to a multi--agency setting. Both of these works deal with OTC trading of derivatives under adverse selection. It should be mentioned that in order to deal with the increased complexity introduced by the enlargement of the set of sellers, additional restrictions had to be imposed on the set of financially feasible products. On the profit maximization side we made use of results of Reny and, more recently, Page Jr. \& Monteiro to show the existence of (mixed--strategies) Nash equilibria. We were unable, however, to obtain a similar result in the case of risk minimization. The non--linear impact of individual contracts on the firms' risk evaluations is incompatible with the machinery developed by the authors previously mentioned. Moreover, even if the game were uniformly payoff secure and (weakly) reciprocal upper semicontinuous, the lack of quasi--convexity would require considering the mixed extension of the game. This brings us to the following
conceptual issue: what is expected risk? Convex risk measures can be (robustly) represented as a worst--case
analysis over a family of probability measures that are absolutely continuous with respect to the reference
one. However, not all of these measures (which can be interpreted as possible distributions of the future states of the World)
are given the same weight. Indeed, they are penalized according to a function that maps the space of probability measures into the
extended Reals. By considering a mixed extension of our risk--minimization game, we implicitly assume that
``Nature'' and the firms behave in qualitatively different ways. In our view this would be inconsistent. Instead, we introduce the notion of socially efficient allocations and prove the existence of such. We believe that an important contribution of this paper is to show that, within our stylized setting, non--regulated, OTC markets cannot be guaranteed to be efficient (in the sense of the welfare theorems). The extension of our general setting to one where agents have heterogenous initial endowments (multi--dimensional agent types) and (we believe more interestingly) to a dynamic framework are left for future research.

\begin{appendix}

\section{Existence of minimizers to $\Psi_1$}\label{app:MP1}
In this appendix we give an overview of the proof of existence of minimizers to problem $\Psi_1$ in Section~\ref{section:Min_risk_fixed_incomes}, which is analogous to the proof of Theorem 2.3 in~\cite{kn:hm}. The main ideas behind the proof are to relax the variance constraint to have a convex minimization problem, and then to show that based on a solution to the latter we can construct a solution to $\Psi_1.$ The steps to be taken are the following:
\begin{enumerate}
\item We fix $(v_1, v_2)\in\C_1\times\C_2$ and relax the variance constraint to  $-\V[X_i(\t)]\le v_i'(\t).$ We observe that
 \be
     \|X_i\|_2^2=v_i(a)-v_i(0)\triangleq k_i^v.
 \ee
Let $K_i$ be the uniform bound on the elements of $\C_i$ (see Prop.~\ref{prop:uniform_bound_v}), then  $\|X_i\|_2^2\le k_i^v\le K_i.$

\item The convexity of $\varrho_i$ implies that the set
 \be
    \X_i^{v, f}\triangleq\big\{X_i\in\X_i\,\mid\, A_i(v_1, v_2, X_i,f)\le\varrho_i(W_i)\big\}
 \ee
is convex. The fact that $\rho_i$ has the Fatou property implies that $\X_i^{v, f}$ is closed.

\item Since the mapping
 \be
     X_i\mapsto\int_{\T_i}X_i(\t)d\t-\int_{\T_0}X_i(\t)f_i(\t)d\t,
 \ee
is a linear, then $X_i\mapsto A_i(v_1, v_2, X_i, f)$ is a convex mapping. The set
 \be
    \X_i^{k_i^v}\triangleq\big\{X\in\X_i\,\mid\,\|X_i\|_2^2\le k_i^v\big\}
 \ee
is a closed,  convex and bounded set, hence the relaxed optimization
problem has a solution $\tilde{X}_i^{v, f},$ as long as $\X_i^{K}\cap\X_i^{v, f}\neq\emptyset.$

\item The variance constraint can be made binding by structuring the
products $\tilde{X}_i^v$ into $X_i^{v, f}=\tilde{X}_i^{v,
f}+\alpha Y.$ Here $Y$ is independent of $\t$ and
$\alpha:\T\to\re$ integrates to zero.
\end{enumerate}

\section{Description of the algorithm used in Sections~\ref{subsec:sim_entropic} and~\ref{subsec:simul_AVAR}}\label{app:Algorithm_risk}

In this appendix we provide a brief description of our numerical algorithm, whose aim is to estimate solutions to the problem
\be
\inf_{(Z, f, v)}\sum_{i=1}^2\Big[\varrho_i\big(W_i-Z_i\int_a^1\sqrt{-v'(\t)}f_i(\t)d\t\big)+\int_a^1(v(\t)-\t\,v'(\t))f_i(\t)d\t\Big]
\ee
subject to:
\begin{itemize}

\item $f_i\ge 0,$ $f_1+f_2=1;$

\item $v$ convex, $v\ge 0,$ $v'\le 0,$ $v(1)=0;$

\item $\E[Z_i]=0,$ $\V[Z_i]=1.$

\item $\varrho_i\big(W_i-Z_i\int_a^1\sqrt{-v'(\t)}f_i(\t)d\t\big)+\int_a^1(v(\t)-\t\,v'(\t))f_i(\t)d\t\leq \varrho_i\big(W_i\big).$ It should be noted that in the examples presented in  Section~\ref{sec:examples}, this constraint was not binding.

\end{itemize}
In order to do so, we set a discretization level $n,$ and we work with the following structures:

\begin{itemize}

\item $f_1(\lambda)\triangleq\sum_{k=0}^{n-1}\lambda_k\i_{[a+k\frac{1-a}{n}, a+(k+1)\frac{1-a}{n}]},\,f_2=1-f_1,\,0\leq\lambda_i\leq 1.$

\item $v'(\cdot)\triangleq\Big(\int_{\cdot}^1 -\frac{d}{d\t}\sqrt{-v'(\t)}d\t+\sqrt{-v'(1)}\Big)^2,$ we shall denote $\tilde{v}\triangleq-\frac{d}{d\t}\sqrt{-v'(\t)},$ and $\gamma\triangleq\sqrt{-v'(1)}\ge 0.$ There is a one--to--one correspondence between $\tilde{v}\ge 0,$ $\gamma\ge 0$ and functions $v$ that satisfy the requirements of being convex, non--increasing and non--negative. This allows for an easier--to--handle description of such functions, instead of having to impose more burdensome, global convexity constraints, which would be necessary in higher dimensions.

\item We go one step further and define $\tilde{v}(\alpha)\triangleq\sum_{k=0}^{n-1}\alpha_k\i_{[a+k\frac{1-a}{n}, a+(k+1)\frac{1-a}{n}]},\,\alpha_i\ge 0.$

\item $Z_i\triangleq(\beta^i_0,\ldots,\beta^i_{d-1})\in\re^d,\,\frac{1}{d}\sum_{k=0}^{d-1}\beta^i_k=0,\,\frac{1}{d}\sum_{k=0}^{d-1}(\beta^i_k)^2\leq 1$. Following the results presented in Appendix~\ref{app:MP1}, we have relaxed $\V[Z_i]=1$ to $\V[Z_i]\leq 1.$

\end{itemize}
In this setting we have that
\be
\varrho_i\big(W_i-Z_i\int_a^1\sqrt{-v'(\t)}f_i(\t)d\t\big)=r_i(\alpha, \beta, \gamma, \lambda).
\ee
and the problem amounts to minimizing the aggregate risk  $\sum_{i=1}^2 r_i(\alpha, \beta, \gamma, \lambda)$ subject to
$r_i(\alpha, \beta, \gamma, \lambda)\leq r_i(0, 0, 0, 0)$ and subject to constraints on $\alpha$, $\beta$, $\gamma$ and $\lambda$ specified earlier. The latter are convex (in fact most of them are linear)in the corresponding variables.
A crucial property of the functions $r_i:\re^{n+2d+1+n}\to\re$ is that for any three fixed entries, they are convex on the remaining variable. Exploiting the latter we use an iterative algorithm that performs a one--step, first--order descent in each direction alternately, such that the aggregate risk decreases in each step. The algorithm repeats this process until a maximal number of iterations has been performed or until there is no significant change in the aggregate risk. At each of the steps we encounter the problem of minimizing a convex function with respect to convex constraints.

\begin{figure}
\begin{center}
  \includegraphics[width=9cm]{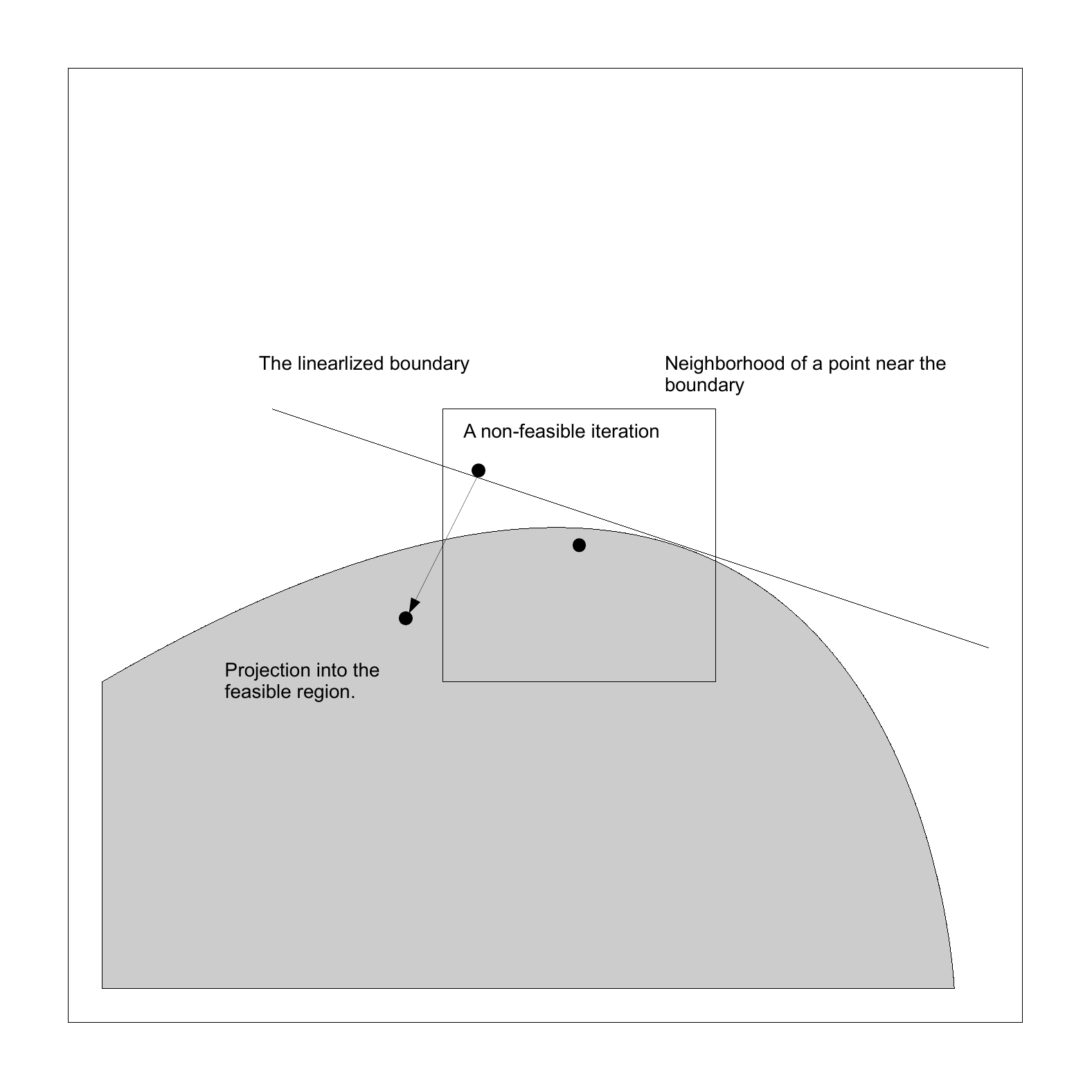}\\
  \caption{Dealing with non--feasible points}\label{figure6}
  \end{center}
\end{figure}

Our coding has been done in Java, and we have used the NetBeans IDE 6.8 environment. In the case of $AV@R,$ we have used the $or124.jar$ package (OR-Objects 1.2.4) to calculate the required gradients. The descent procedures are all based on a local linearization of the function to be minimized, as well as of the constraints. Since all of these objects are convex, they can be locally, well approximated by the corresponding subgradients (which are calculated using the chain rule). The linearized function is minimized, subject to the linearized constraints, on a cubic neighborhood of the current point. This reduces to a linear optimization problem. Again the linear optimization package included in $or124.jar$ is used to obtain a minimizer. If this point does not satisfy the constraints (we carry the linearization error), a correction procedure is performed to obtain a feasible point close to the prior one. This procedure essentially performs a line search in the direction opposite to the direction of the subgradient of the constraint function at the point in hand (See Figure~\ref{figure6}). Finally, it is verified whether at this feasible point the value of the objective function has decreased. Otherwise, the procedure is repeated starting with a cube whose size length is half of the original one.

\section{Convex risk measures on $\Lp^{2}$.}\label{app:RM}

In this appendix we recall some properties and representation
results for risk measures on $\Lp^2(\O, \Prob).$ We refer the reader
to~\cite{kn:chl} and to section 4.3 in~\cite{kn:fs} for detailed discussions on this topic.

\begin{definition}
\begin{rmenumerate}
\item A {\sl{monetary measure of risk}} on $\Lp^2$ is a function
$\varrho:\Lp^{2}(\O, \Prob)\to\re\cup\{+\infty\}$ such that for all $X,Y \in\Lp^{2}$ the
following conditions are satisfied:
\begin{itemize}
\item Monotonicity: if $X\le Y$ then $\varrho(X)\ge\varrho(Y)$.

\item Cash Invariance: if $m\in\re$ then $\varrho(X+m)=\varrho(X)-m$.
\end{itemize}
\item A risk measure is called {\sl coherent} if it is convex and homogeneous of degree 1, i.e., if the following two conditions hold:
\begin{itemize}
\item Convexity: for all $\lambda \in [0,1]$ and all positions
$X,Y \in\Lp^{2}$:
\be
    \varrho(\lambda X+(1-\lambda)Y)\le\lambda\varrho(X)+(1-\lambda)\varrho(Y)
\ee

\item Positive Homogeneity: For all $\lambda \geq 1$
\be
    \varrho(\lambda X)=\lambda\varrho(X).
\ee
\end{itemize}
\item The risk measure is called law invariant, if
\be
    \varrho(X)=\varrho(Y)
\ee
for any two random variables $X$ and $Y$ which have the same law.
\item The risk measure $\varrho$ on $\Lp^{2}$ has the {\sl Fatou
property} if for any sequence of random variables $X_1,
X_2,\ldots$ that converges in norm to a random variable $X$ we
have
\be
    \varrho(X)\le\liminf_{n\to\infty} \varrho(X_n).
\ee
\end{rmenumerate}
\end{definition}
Since a risk measure $\varrho$ that has the Fatou property is a l.s.c. and proper convex mapping from $\Lp^{2}(\O, \Prob)$ into $\re\cup\{+\infty\},$
we may represent it via a Legendre-Fenchel transform. Namely, if we define $\mm_1(\Prob)$ to be the set of probability measures on $\O$ that are absolutely continuous (w.r.t $\Prob$) and
\be
\mm_1^2(\Prob)\triangleq\Big\{ Q\in \mm_1(\Prob)\,\big|\, \dfrac{dQ}{d\Prob}\in\Lp^2\Big\},
\ee
then given $\varrho$ there exists a penalty function $\alpha:\Lp^{2}(\O, \Prob)\to\re\cup\{+\infty\}$ such that
\be
\varrho(X)=\sup_{Q\in \mm_1^2(\Prob)}\big\{\E_{\Prob}[-X]-\alpha(Q)\big\}.
\ee
Moreover, if for $Q\in \mm_1^2(\Prob)$ we write $Y=dQ/d\Prob,$ then the above expression can be written as
\be
\varrho(X)=\sup_{\|Y\|_2=1}\big\{-\langle X, Y\rangle-\beta(Y)\big\},
\ee
where $\langle \cdot, \cdot\rangle$ is the canonical inner product in $\Lp^2(\O, \Prob),$ and $\beta(Y)=\alpha(Q).$
\end{appendix}


\end{document}